\documentclass[ALICE,manyauthors]{cernphprep}
\usepackage[comma,square,numbers,sort&compress]{natbib}

\usepackage{lineno}
\usepackage{xspace}
\usepackage{hyperref}
\usepackage{upgreek}

\usepackage[T1]{fontenc}
\usepackage{orcidlink}

\begin{document}

\definecolor{light-gray}{gray}{0.8}

\newcommand{\pp}{\ensuremath{\rm pp}\xspace}
\newcommand{\pPb}{p--Pb\xspace}
\newcommand{\PbPb}{Pb--Pb\xspace}
\newcommand{\AuAu}{Au--Au\xspace}
\newcommand{\ppbar}{\ensuremath{\rm p\bar{p}}\xspace}
\newcommand{\epluseminus}{\ensuremath{\rm e^{+}e^{-}}\xspace}
\newcommand{\ep}{\ensuremath{\rm ep}\xspace}

\newcommand{\MeVc}{\ensuremath{{\rm MeV/}c}\xspace}
\newcommand{\MeVcsq}{\ensuremath{{\rm MeV/}c^{2}}\xspace}
\newcommand{\GeVc}{\ensuremath{{\rm GeV/}c}\xspace}
\newcommand{\GeVcsq}{\ensuremath{{\rm GeV/}c^{2}}\xspace}
\newcommand{\eV}{\ensuremath{\rm eV}\xspace}
\newcommand{\keV}{\ensuremath{\rm keV}\xspace}
\newcommand{\MeV}{\ensuremath{\rm MeV}\xspace}
\newcommand{\GeV}{\ensuremath{\rm GeV}\xspace}
\newcommand{\TeV}{\ensuremath{\rm TeV}\xspace}
\newcommand{\fm}{\ensuremath{\rm fm}\xspace}
\newcommand{\mm}{\ensuremath{\rm mm}\xspace}
\newcommand{\cm}{\ensuremath{\rm cm}\xspace}
\newcommand{\m}{\ensuremath{\rm m}\xspace}
\newcommand{\mum}{\ensuremath{\mu{\rm m}}\xspace}
\newcommand{\s}{\ensuremath{\rm s}\xspace}
\renewcommand{\d}{\ensuremath{\rm d}\xspace}
\newcommand{\ns}{\ensuremath{\rm ns}\xspace}
\newcommand{\ps}{\ensuremath{\rm ps}\xspace}
\newcommand{\mrad}{\ensuremath{\rm mrad}\xspace}
\newcommand{\mb}{\ensuremath{\rm mb}\xspace}
\newcommand{\ctau}{\ensuremath{{\rm c}\tau}\xspace}
\newcommand{\ctauapprox}[1]{\ctau $\approx$ #1 \mum}

\newcommand{\sqrts}{\ensuremath{\sqrt{s}}\xspace}
\newcommand{\compp}{{\ensuremath{\sqrt{s}} = 5.02 TeV}\xspace}
\newcommand{\sqrtsNN}{\ensuremath{\sqrt{s_{\rm NN}}}\xspace}
\newcommand{\pt}{\ensuremath{{\it p}_{\rm T}}\xspace}
\newcommand{\ptD}{\ensuremath{{\it p}_{\rm T, D^{0}}}\xspace}
\newcommand{\ptjet}{\ensuremath{{\it p}_{\rm T}^{\rm ch~jet}}\xspace}
\newcommand{\come}[1]{{\ensuremath{\sqrt{s}} =  #1 TeV}\xspace}
\newcommand{\comeNN}[1]{{\sqrtsNN  =  #1 TeV}\xspace}

\newcommand{\nch}{\ensuremath{N_{\rm ch}}\xspace}
\newcommand{\ntracklets}{\ensuremath{N_{\rm tracklets}}\xspace}
\newcommand{\ntrackletscorr}{\ensuremath{N^{\rm corr}_{\rm tracklets}}\xspace}
\newcommand{\dnchdeta}{\ensuremath{{\rm d}N_{\rm ch}/{\rm d}\eta}\xspace}
\newcommand{\dnchdpt}{\ensuremath{{\rm d}N_{\rm ch}/{\rm d}\pt}\xspace}
\newcommand{\dnchdy}{\ensuremath{{\rm d}N_{\rm ch}/{\rm d}y}\xspace}
\newcommand{\dntrackletsdeta}{\ensuremath{{\rm d}N_{\rm tracklets}/{\rm d}\eta}\xspace}
\newcommand{\avgntracklets}{\ensuremath{<N_{\rm tracklets}>}\xspace}
\newcommand{\avgntrackletscorr}{\ensuremath{<N^{\rm corr}_{\rm tracklets}>}\xspace}

\newcommand{\ptmore}[1]{{\pt $>$ #1 \GeVc}\xspace}
\newcommand{\ptless}[1]{{\pt $<$ #1 \GeVc}\xspace}
\newcommand{\ptrange}[2]{{#1 $<$ \pt $<$ #2  \GeVc}\xspace}
\newcommand{\ptDrange}[2]{{#1 $<$ \ptD $<$ #2  \GeVc}\xspace}
\newcommand{\jetptrange}[2]{{#1 $\leq$ \ptjet $<$ #2  \GeVc}\xspace}
\newcommand{\etaless}[1]{\ensuremath{\left|\eta\right| < #1}\xspace}
\newcommand{\etarange}[2]{\ensuremath{#1 < \eta < #2}\xspace}
\newcommand{\yless}[1]{\ensuremath{\left|y\right| < #1}\xspace}
\newcommand{\yfid}{\ensuremath{y_{\rm fid}(\pt)}\xspace}
\newcommand{\yfidcut}{\ensuremath{\left|y\right| < y_{\rm fid}(\pt)}\xspace}
\newcommand{\dnchdetaless}[1]{\ensuremath{{\rm d}N_{\rm ch}/{\rm d}\eta|_{\etaless{#1}}}\xspace}
\newcommand{\zvertex}{\ensuremath{z_{\rm vtx}}\xspace}
\newcommand{\zvertexless}[1]{\ensuremath{\left|z_{\rm vtx}\right| < #1\ \cm}\xspace}
\newcommand{\zvertexout}[1]{\ensuremath{\left|z_{\rm vtx}\right| > #1\ \cm}\xspace}

\newcommand{\slfrac}[2]{\ensuremath{\left.#1\right/#2}\xspace}
\newcommand{\dedx}{\ensuremath{{\rm d}E/{\rm d}x}\xspace}
\newcommand{\asymmerr}[2]{\ensuremath{^{+#1}_{-#2}}\xspace}
\newcommand{\average}[1]{\ensuremath{\langle #1 \rangle}\xspace}
\newcommand{\nluminosity}[2]{\ensuremath{L_\text{int} = {#1}\pm{#2}\ \text{nb}^{-1}}\xspace}
\newcommand{\luminosity}{\ensuremath{L_\text{int}}\xspace}

\newcommand{\sphero}{\ensuremath{{S}_{\textit O}}\xspace}
\newcommand{\Dzero}{\ensuremath{{\rm D}^{0}}\xspace}
\newcommand{\Dzerobar}{\ensuremath{\overline{{\rm D}^{0}}}\xspace}
\newcommand{\DzeroNS}{\ensuremath{{\rm D}^{0}}}       %NS = No space
\newcommand{\Dstarplus}{\ensuremath{{\rm D}^{*+}}\xspace}
\newcommand{\DstarplusNS}{\ensuremath{{\rm D}^{*+}}}       %NS = No space
\newcommand{\Dplus}{\ensuremath{{\rm D}^{+}}\xspace}
\newcommand{\DplusNS}{\ensuremath{{\rm D}^{+}}}       %NS = No space
\newcommand{\Piplus}{\ensuremath{{\pi}^{+}}\xspace}
\newcommand{\PiplusNS}{\ensuremath{{\pi}^{+}}}        %NS = No space
\newcommand{\Piminus}{\ensuremath{{\pi}^{-}}\xspace}
\newcommand{\PiminusNS}{\ensuremath{{\pi}^{-}}}
\newcommand{\sPi}{\ensuremath{{\pi}}\xspace}
\newcommand{\Kplus}{\ensuremath{{\rm K}^{+}}\xspace}
\newcommand{\KplusNS}{\ensuremath{{\rm K}^{+}}}
\newcommand{\Kminus}{\ensuremath{{\rm K}^{-}}\xspace}
\newcommand{\KminusNS}{\ensuremath{{\rm K}^{-}}}
\newcommand{\sProton}{\ensuremath{\rm p}\xspace}
\newcommand{\pProtonNS}{\ensuremath{\rm p}}
\newcommand{\apProton}{\ensuremath{\overline{\rm p}}\xspace}
\newcommand{\apProtonNS}{\ensuremath{\overline{\rm p}}}
\newcommand{\sKzero}{\ensuremath{2{\rm K}^{0}_{S}}\xspace}
\newcommand{\pKzero}{\ensuremath{{\rm K}^{0}_{S}}\xspace}
\newcommand{\sXi}{\ensuremath{\Xi}\xspace}
\newcommand{\pXi}{\ensuremath{\Xi^{-}}\xspace}
\newcommand{\pXiNS}{\ensuremath{\Xi^{-}}}
\newcommand{\apXi}{\ensuremath{\overline{\Xi}^{+}}\xspace}
\newcommand{\apXiNS}{\ensuremath{\overline{\Xi}^{+}}}
\newcommand{\Jpsi}{\ensuremath{{\rm J}/\psi}\xspace}

\newcommand{\DtoKpi}{\ensuremath{{\Dzero}\to{\KminusNS}{\PiplusNS}}\xspace} 
\newcommand{\DbartoKpi}{\ensuremath{{\Dzerobar}\to{\KplusNS}{\PiminusNS}}\xspace}   
\newcommand{\DtoKpipi}{\ensuremath{{\Dplus}\to{\KminusNS}{\PiplusNS}{\PiplusNS}}\xspace}   
\newcommand{\DstartoDpi}{\ensuremath{{\Dstarplus}\to{\DzeroNS}{\PiplusNS}}\xspace}   

\newcommand{\brDtoKpi}{\ensuremath{{\rm 3.93} \pm {\rm 0.04}\%}\xspace}   
\newcommand{\brDtoKpipi}{\ensuremath{{\rm 9.46} \pm {\rm 0.24}\%}\xspace}   

\newcommand{\error}[1]{{\color{red}\bf \textsc{Wrong:} #1}}
\newcommand{\warning}[1]{{\color{red}\bf \textsc{Warning:} #1}}
\newcommand{\comments}[1]{{\color{blue}\bf \textsc{Comment:} #1}}
\newcommand{\correct}[2]{{\color{red}\sout{#1}}{\color{blue}\xspace#2}}
\newcommand{\cmnt}[1]{}
\newcommand{\fake}[1]{{\color{red}\bf#1}}

\newcommand{\alphas}{\ensuremath{\alpha_{\rm S}}\xspace}
\newcommand{\SPS}{\ensuremath{\rm Sp\bar{p}S}\xspace}
\newcommand{\Raa}{\ensuremath{R_{\rm AA}}\xspace}
\newcommand{\Rppb}{\ensuremath{R_{\rm pPb}}\xspace}
\newcommand{\Rpa}{\ensuremath{R_{\rm pA}}\xspace}

\newcommand{\Rdaxis}{\ensuremath{\Delta R_{\rm D, jet/axis}}\xspace}

\newcommand{\deltaR}[2]{{$\Delta R_{\rm #1 - #2}$}\xspace}

\newcommand{\R}{\ensuremath{\Delta R_{\rm D, jet}}\xspace}

\newcommand{\Raxis}{\ensuremath{\Delta R_{\rm axis}}\xspace}

\newcommand{\Rplain}{\ensuremath{\Delta R_{\rm axis}}\xspace}

\newcommand{\zcut}{\ensuremath{z_{\rm cut}}\xspace}

\newcommand{\jetpT}{\ensuremath{p_{\rm T}^{\rm ch~jet}}\xspace}

\newcommand{\dzeropT}{\ensuremath{p_{\rm T}^{\rm D^0}}\xspace}

\newcommand{\detjetpT}{\ensuremath{p_{\rm T,~det}^{\rm ch~jet}}\xspace}

\newcommand{\truthjetpT}{\ensuremath{p_{\rm T,~truth}^{\rm ch~jet}}\xspace}

\newcommand{\detdeltaR}{\ensuremath{\Delta R_{\rm det}}\xspace}

\newcommand{\truthdeltaR}{\ensuremath{\Delta R_{\rm truth}}\xspace}

%\linenumbers

%%%%%%%%%%%%%%%  Title page %%%%%%%%%%%%%%%%%%%%%%%%
\begin{titlepage}
% the dates below correspond to CERN approval
% please don't touch: EB chairs will take care
\PHyear{2025}       % required, will be obtained from CERN
\PHnumber{083}      % required, will be obtained from CERN
\PHdate{01 April}  % required, will be obtained from CERN
%%%%%%%%%%%%%%%%%%%%%%%%%%%%%%%%%%%%%%%%%%%%%%%%%%%%

%%% Put your own title + short title here:
\title{D$^{\bf 0}$-meson-tagged jet axes difference in proton--proton collisions at $\mathbf{\sqrt{\textit{s}} = 5.02}$ TeV}
\ShortTitle{D$^{0}$-meson-tagged jet axes difference in pp}   % appears on left page headers

%%% Do not change the next lines
\Collaboration{ALICE Collaboration\thanks{See Appendix~\ref{app:collab} for the list of collaboration members}}
\ShortAuthor{ALICE Collaboration} % appears on right page headers, do not change

\begin{abstract}
Heavy-flavor quarks produced in proton--proton (pp) collisions provide a unique opportunity to investigate the evolution of quark-initiated parton showers from initial hard scatterings to final-state hadrons. By examining jets that contain heavy-flavor hadrons, this study explores the effects of both perturbative and non-perturbative QCD on jet formation and structure. The angular differences between various jet axes, $\Delta R_{\rm axis}$, offer insight into the radiation patterns and fragmentation of charm quarks. The first measurement of D$^{0}$-tagged jet axes differences in pp collisions at $\sqrt{s}=5.02$ TeV by the ALICE experiment at the LHC is presented for jets with transverse momentum $p_{\rm T}^{\rm ch~jet} \geq 10$ ${\rm GeV}/c$ and D$^0$ mesons with $p_{\rm T}^{\rm D^{0}} \geq 5$ ${\rm GeV}/c$. In this D$^0$-meson-tagged jet measurement, three jet axis definitions, each with different sensitivities to soft, wide-angle radiation, are used: the Standard axis, Soft Drop groomed axis, and Winner-Takes-All axis. Measurements of the radial distributions of D$^0$ mesons with respect to the jet axes, $\Delta R_{\mathrm{axis-D^0}}$, are reported, along with the angle, $\Delta R_{\mathrm{axis}}$, between the three jet axes. The D$^{0}$ meson emerges as the leading particle in these jets, closely aligning with the Winner-Takes-All axis and diverging from the Standard jet axis. The results also examine how varying the sensitivity to soft radiation with grooming influences the orientation of the Soft Drop jet axis, and uncover that charm-jet structure is more likely to survive grooming when the Soft Drop axis is further from the D$^{0}$ direction, providing further evidence of the dead-cone effect recently measured by ALICE.

\end{abstract}
\end{titlepage}

\setcounter{page}{2} %please do not remove this line            

\section{Introduction}\label{sec:introduction}

High-energy collisions at particle colliders produce energetic quarks and gluons, which are scattered at wide angles relative to the beam direction. These partons emit radiation and split into lower-energy partons, forming a parton shower, until their energy drops below the quantum chromodynamics (QCD) confinement scale. At this point, the parton shower transitions into collimated sprays of color-neutral hadrons, known as jets, which serve as powerful probes for studying QCD~\cite{event_generators_LHCphysics_2011,ALICEjetprediction,ATLASjetprediction,CMSjetprediction}. A detailed understanding of parton showers within perturbative QCD (pQCD) is of high interest and new measurements provide valuable input to jet fragmentation and hadronization models, especially when using observables that maintain control over non-perturbative effects~\cite{Cruz_Torres_2022,wenqing_EEC_pp,groom_jetangularities_pp_2022,atlas_jetfragmentation_2019,atlas_softdrop_2020,atlas_lund_2020,cms_mult_shape2012,star_groomjet_pp_2020}. Measurements in pp collisions serve as a reference for future measurements in heavy-ion collisions, where the radiation pattern, and therefore the jet substructure, is modified by a deconfined state of quarks and gluons called the quark--gluon plasma (QGP). Measurements of these modifications probe this phase of QCD matter~\cite{Busza_2018, ALICE:2008, pp_pPb_jetstructure, jetaxes_PbPb, minjung_jpsi_2023, nsubjetti_PbPb_2021, subjet_frag_pp_PbPb2023, decluster_pp_pPb_2020, atlas_suppression_PbPb_2023, cms_jetfrag_pp_PbPb_2012, lhcb_jpsi_2017, star_jetstructure_AuAu_2022, formation_partonicmatter_phenix_2005}.

Heavy-flavor quarks, namely charm and beauty, provide a unique opportunity to trace the evolution of quark-initiated parton showers, from their production in the initial hard scatterings to the formation of final-state hadrons~\cite{dead_cone_2022,d0_prod_5and13TeV_2023,bjetcjet_pp_viaPromptNonprompt_2021,bayrontomeson_pp_2023,cms_lambda_pp_PbPb_2020,charm_eec}. Due to the finite quark mass, heavy-quark production can be computed using pQCD by introducing a hard scale that effectively suppresses non-perturbative effects and ensures a stable perturbative expansion~\cite{prompt_charm_7TeV_2023, charm_prod_2TeV_2012, charm_production_7TeV_2012}. Quark-initiated showers are expected to have a narrower fragmentation profile than those initiated by gluons, due to a difference in the Casimir color factors between quarks and gluons~\cite{charm_eec, casimir}. The fragmentation profile of parton showers initiated by a heavy quark are also generally harder than those initiated by a light quark or a gluon because of an additional mass-dependent suppression of the gluon radiation emitted by the heavy quark, colloquially known as the ``dead-cone effect"~\cite{deadcone_theory_1991}. This suppression was directly observed by the ALICE experiment~\cite{dead_cone_2022} and is known to be more significant for low-momentum jets because of the inverse relation of the dead-cone angle with the energy of the radiating quark, $\theta_{\rm d.c.} \approx m_{\rm q}/E_{\rm radiator}$. Comparisons between jets containing a heavy quark and inclusive jet samples reveal the influence of Casimir color factors and the quark mass on the fragmentation and subsequent hadronization processes within the QCD shower.

This work leverages a novel probe of soft, wide-angle radiation to study the effects of perturbative and non-perturbative QCD processes on jet fragmentation and radiation patterns. The jet axes difference, $\Delta R_{\rm axis}$, is the angular distance between two different definitions for the axis of the jet in pseudorapidity, $\eta$, and azimuth, $\varphi$, given by 
\begin{equation}
    \Delta R_{\rm axis} = \sqrt{(\Delta\eta)^{2}+(\Delta\varphi)^{2}}.
    \label{eq:deltaR}
\end{equation}

The jet axes difference is an infrared-collinear safe observable even at Next-to-Leading Order (NLO)~\cite{Cal:2019gxa}. It is calculable in the small-angle limit within Soft Collinear Effective Theory (SCET), but measurements are additionally sensitive to the non-perturbative physics not included in this perturbative framework. Jet axes differences have been studied for inclusive, gluon-dominated jets in ALICE~\cite{Cruz_Torres_2022}, but have not been previously studied for heavy-flavor jets. 

This work presents the first measurement of this observable for charm-tagged jets, selected by requiring the jets to contain a \Dzero meson or its charge conjugate. The three jet axes chosen for study in this analysis are the Standard, Soft Drop, and Winner-Takes-All axes (labeled in the figures and tables as STD, SD, and WTA, respectively). Comparisons of the data to several event generators are discussed. To probe flavor dependencies in QCD showers, predictions from PYTHIA 8 of inclusive, gluon-initiated, and light-flavor-initiated jets are compared to the charm-initiated results.

Jets are reconstructed using the anti-$k_{\rm T}$ algorithm with resolution parameter $R=0.4$~\cite{antikT_2008} and $E$-scheme recombination~\cite{fastjet_2012}. The $E$-scheme recombination strategy combines the tracks of the jet, and defines the four-momenta of the jet as the sum of the constituents' four-momenta.
This also defines the Standard jet axis. The jet constituents can be reclustered using the Cambridge--Aachen (angular ordering) algorithm with the same resolution parameter~\cite{cambridge_aachen1}. From this, the Soft Drop and Winner-Takes-All axes can be calculated. 

The Soft Drop axis is determined by starting at the root splitting of the reclustered jet, following the harder branch iteratively through the clustering tree, and apply the Soft Drop condition at each splitting,
\begin{equation}
    \frac{\min(p_{\rm T,1}, p_{\rm T,2})}{p_{\rm T,1} + p_{\rm T,2}} > z_{\text{cut}} \left( \frac{\Delta R_{1,2}}{R} \right)^{\beta},
    \label{eq:softdropcondition}
\end{equation}
where $p_{\rm T,1}$ and $p_{\rm T,2}$ are the transverse momenta of the two prongs and $\Delta R_{1,2}$ is their angular difference in the rapidity-azimuth plane. The tunable parameters, \( z_{\text{cut}} \) and \( \beta \), control the grooming intensity and determine if the splitting is removed. Specifically, \( z_{\text{cut}} \) sets the threshold for asymmetry in transverse momentum between split branches, while \( \beta \) adjusts the weight given to their angular separation~\cite{softdrop_Larkoski}. The grooming procedure ends when a splitting that satisfies the Soft Drop condition is found, with its parent branch defining the groomed jet axis. If no splitting passes, the jet is excluded from the Soft Drop sample.  

The Winner-Takes-All scheme also operates on the reclustered jet. At each splitting, the direction of the hardest prong is selected with a transverse momentum equal to the \pt sum of the two sub-prongs~\cite{wta_Larkoski}. 

\begin{figure}[!htb]
\centering
\includegraphics[scale=0.3]{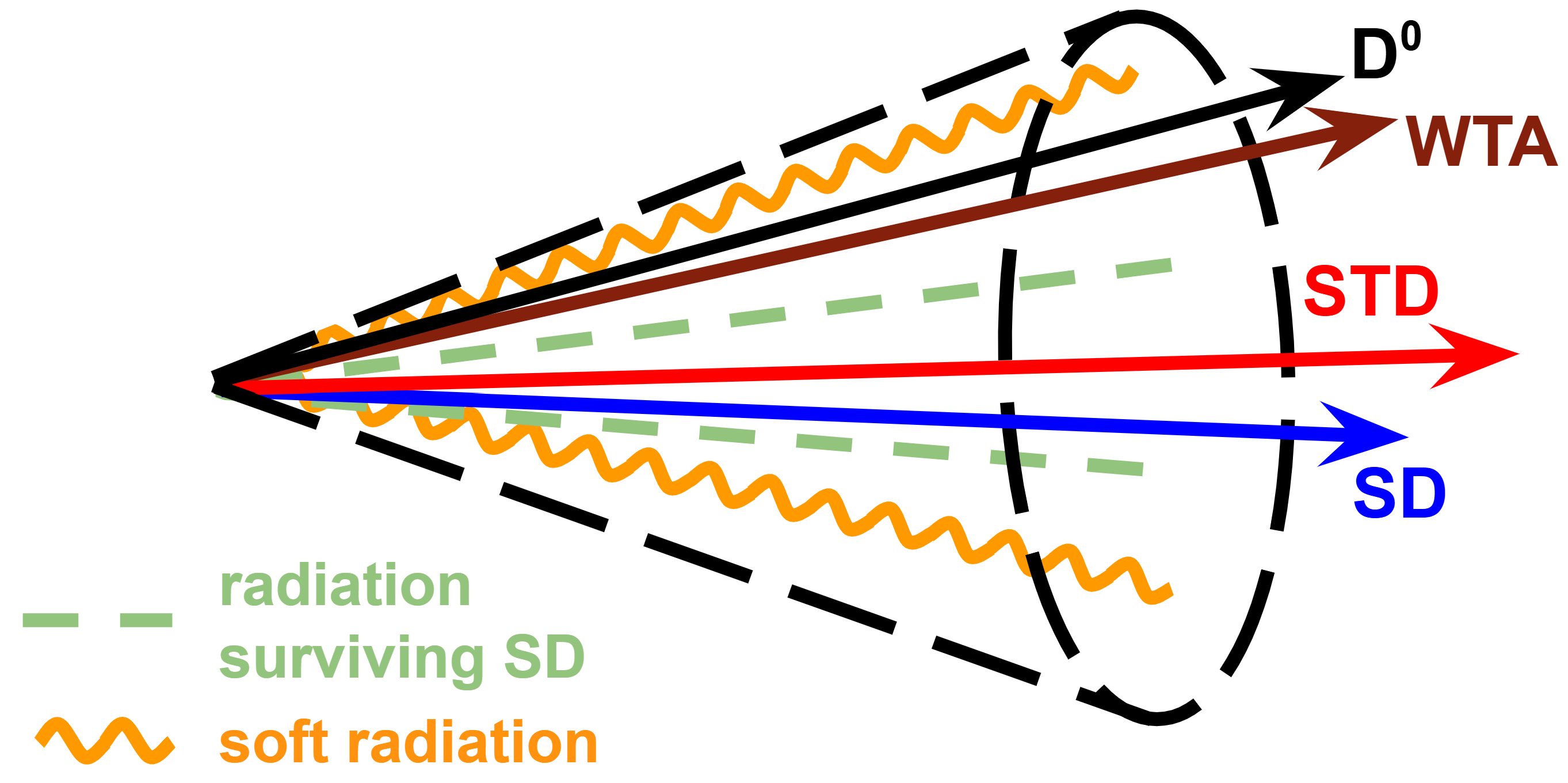}
\hspace{0.5cm}
\includegraphics[scale=0.6]{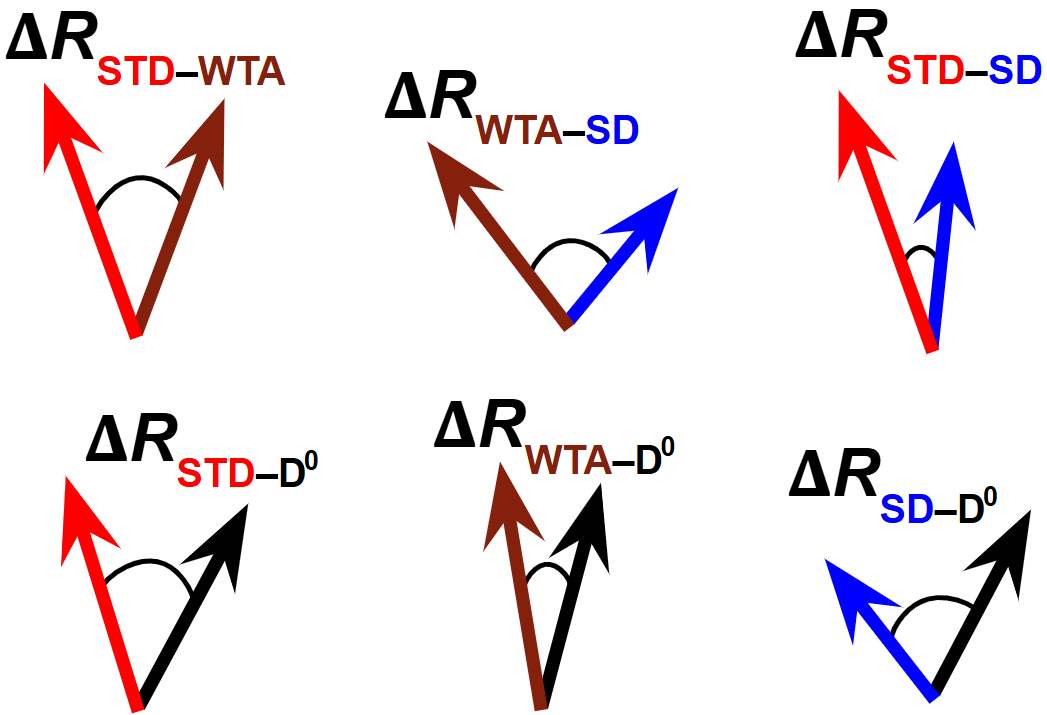}
\caption{A representation of the different jet axes; Standard (STD), Soft Drop (SD), and Winner-Takes-All (WTA). The axis of the jet containing the \Dzero meson (shown in solid black) and all gluonic radiation is referred to as the Standard axis. The Winner-Takes-All axis is also determined from this initial jet sample but aligns with the hardest subjet at each clustering step. Grooming away softer radiation leaves a jet with the Soft Drop axis, defined by the remaining higher-momentum particles. Representations for the six \Raxis observables are shown in the right of the figure.}
\label{fig:jetaxesdiff}
\end{figure}

Six \Rplain observables are measured in this paper, summarized in Fig.~\ref{fig:jetaxesdiff}. The jet axes vary in their sensitivity to soft, non-perturbative radiation, with the Standard axis being the most responsive to these effects. Grooming removes radiation according to the intensity of the grooming parameters, $\zcut$ and $\beta$, making the Soft Drop jet less sensitive to soft radiation than an ungroomed jet. The Winner-Takes-All axis is insensitive to soft radiation at leading order. Comparing the angles between the three axes (\deltaR{STD}{WTA}, \deltaR{WTA}{SD}, and \deltaR{STD}{SD}) tunes the sensitivity to these non-perturbative effects and investigate the radiation pattern inside the reconstructed jets. The difference between the \Dzero direction and the three jet axes (\deltaR{STD}{\Dzero}, \deltaR{WTA}{\Dzero}, and \deltaR{SD}{\Dzero}) allows for studies of the flavor dependence of the fragmentation process~\cite{Cruz_Torres_2022,softdrop_Larkoski,wta_Larkoski, Cacciari_charm_frag}.

This paper is organized as follows:
Sec.~\ref{sec:setupanddatasample} describes the experimental setup of the ALICE detector and the data used in this analysis. Section~\ref{sec:analysismethods} delineates the steps of the analysis and Sec.~\ref{sec:systematics} discusses the related systematic uncertainties. Section~\ref{sec:results} presents the results and discussion, and Sec.~\ref{sec:conclusion} summarizes the conclusions.

\section{The ALICE detector and data sample}\label{sec:setupanddatasample}

This work analyzes pp collision data at $\sqrt{s} = 5.02$ TeV recorded in 2017 by the ALICE experiment at the LHC~\cite{ALICE:2008}. ALICE has a pseudorapidity acceptance of $|\eta|<0.9$ in the central barrel and full azimuthal coverage. It has excellent particle identification (PID) and precise tracking and spatial resolution for charged particles and decay vertices, which are required for reconstructing the \Dzero mesons in this analysis. A detailed description of the experimental setup and performance at ALICE can be found at Ref.~\cite{ALICE_performance_2014}.

The reconstruction of the tracks of charged particles utilized in this analysis employs three detectors: the Inner Tracking System (ITS), the Time Projection Chamber (TPC), and the Time-of-Flight detector (TOF). All three detectors are located in the central barrel of the experiment. Charged particles are detected in the ITS, a six-layer cylindrical silicon detector, and the TPC, a gaseous drift chamber. Reconstruction of low-momentum particles down to $0.15 \ \GeVc$ in the TPC is possible because of the low magnetic field at ALICE (0.5 T) and the low material budget. Particles in this analysis are required to have $\pt > 0.15 \ \GeVc$ and a pseudorapidity $|\eta|<0.9$. The transverse-momentum resolution $\sigma(p_{\rm T})/p_{\rm T}$ increases linearly from $\approx 1\%$ at $p_{\rm T}=1 \ \GeVc$ to $\approx4\%$ at $p_{\rm T}=50 \ \GeVc$. The tracking efficiency rapidly increases from $\approx 60\%$ at $p_{\rm T}=0.15 \ \GeVc$ to $\approx 80\%$ at $p_{\rm T}=1 \ \GeVc$, and remains above $\approx 75\%$ at higher $p_{\rm T}$.

The spatial precision of track reconstruction in the ITS and TPC allows for precise identification of charm hadrons near the interaction point and, consequently, the charm-hadron decay vertices. The mean proper decay length of the \Dzero meson is $c \tau \approx$ 123.01 $\upmu \rm m $~\cite{pdg_2024}, therefore the secondary decay vertex is displaced, typically, by a few hundred $\upmu \rm m$ from the primary collision vertex. A distance of closest approach selection between decay daughters is used to enhance the signal-to-background ratio for \Dzero decay products.

Charged particles are identified using the time of flight from the collision vertex, measured by the TOF, and the specific energy loss, ${\rm d}E/{\rm d}x$, in the TPC. The TPC provides the largest kaon/pion separation at low \pt ($\pt \lesssim 0.7$ \GeVc) and, because of the relativistic rise in ${\rm d}E/{\rm d}x$, still achieves good separation at high \pt ($\pt \gtrsim 2$ \GeVc)~\cite{ALICE_performance_2014}. The TOF is particularly effective in separating the kaon and pion in the intermediate \pt region, offering crucial PID capabilities for the \Dzero decay daughters in the kinematic range where TPC PID is unable to discriminate pions and kaons. 

The data samples analyzed in this work consist of 870 million minimum bias events from pp collisions at $\sqrt{s}=5.02$ TeV, corresponding to an integrated luminosity of $(18.0 \pm 0.4)$ nb$^{-1}$~\cite{wenqing_EEC_pp}. Minimum-bias events are triggered by requiring a signal in both V0 scintillator detectors~\cite{V0_info}, which cover full acceptance in azimuth and the pseudorapidity ranges $2.8 < \eta < 5.1$ (V0A) and $-3.7 < \eta < -1.7$ (V0C). Events are also required to have a primary vertex within 10 cm of the nominal interaction point to provide uniform pseudorapidity acceptance. The Monte Carlo (MC) samples, used in corrections throughout the analysis, are charm-enriched and were produced using the PYTHIA 8.2 event generator with the Monash 2013 tune~\cite{pythia_monash_2014} transported through a GEANT 3~\cite{geant3_Brun:1119728} simulation of the ALICE detector. The charm content was enhanced by requesting a c$\rm \bar{c}$ in 50\% of the events and a b$\rm \bar{b}$ in the remaining half. The MC samples were anchored to the detector conditions during the data collection run-by-run. 
\section{Analysis methods}
\label{sec:analysismethods}
\subsection{D$^{\bf 0}$ meson and jet reconstruction}

Jets initiated by charm quarks are identified by the presence of a prompt \Dzero meson among their constituents. Prior to jet finding, the \Dzero-meson candidates are reconstructed from their hadronic decay mode, \Dzero $\rightarrow \rm K^- + \uppi^+$ (and its charge conjugate), which has a branching ratio of $(3.947 \pm 0.030)\%$~\cite{pdg_2024}. The \Dzero-meson candidates undergo topological selections based on the displacement of the decay vertex, as well as particle-identification selections.  Details of the tracking and \Dzero-meson selections can be found in Ref.~\cite{d0_prod_5and13TeV_2023}, where the same selections are used. The selection criteria significantly suppress the combinatorial background from hadrons that do not originate from \Dzero decays. The topological selections are tuned to optimize the signal-to-background ratio while suppressing the background from the beauty-initiated, non-prompt \Dzero, which is typically enhanced by these selections due to the larger displacement from the collision point of the non-prompt \Dzero compared to the prompt. After the \Dzero candidates are reconstructed, the 4-momenta of the daughters of each candidate are replaced by their vector sum (representing the \Dzero 4-momentum), such that the charm quark remains traceable through the splitting tree after jet reconstruction is performed. This mitigates against cases where the angle between the decay daughters is larger than the jet diameter, which would degrade the jet-energy resolution, and allows for efficient tagging of charm jets by the presence of a \Dzero candidate among their constituents.

The four-momenta of each \Dzero candidate and all other reconstructed tracks are used as input to the anti-$k_{\rm T}$ jet finding algorithm implemented in FastJet~\cite{fastjet_2012}. A minimum \Dzero transverse momentum, $\dzeropT \geq 5 \ \GeVc$, is required, with a corresponding \Dzero acceptance in rapidity of $|y_{\rm D^0}| \leq 0.8$. This analysis concentrates on jets with transverse momenta \jetptrange{10}{20}, but also extends to \jetptrange{20}{50} for the study of the difference between the Winner-Takes-All and \Dzero axes shown in Sec.~\ref{secsec:wta-d}. The Standard jet axis was required to be within $|\eta_{\rm jet}| < 0.9-R$, where $R=0.4$ is fixed for this analysis, to guarantee that the full jet cone is contained within the TPC acceptance. The Standard, Soft Drop, and Winner-Takes-All axes, along with the \Dzero meson 
direction, are determined for each \Dzero-tagged jet.

\subsection{Raw-yield and shape extraction procedure}\label{subsec:yieldextraction}
\begin{figure}[!htb]
\centering
\includegraphics[scale=0.37]{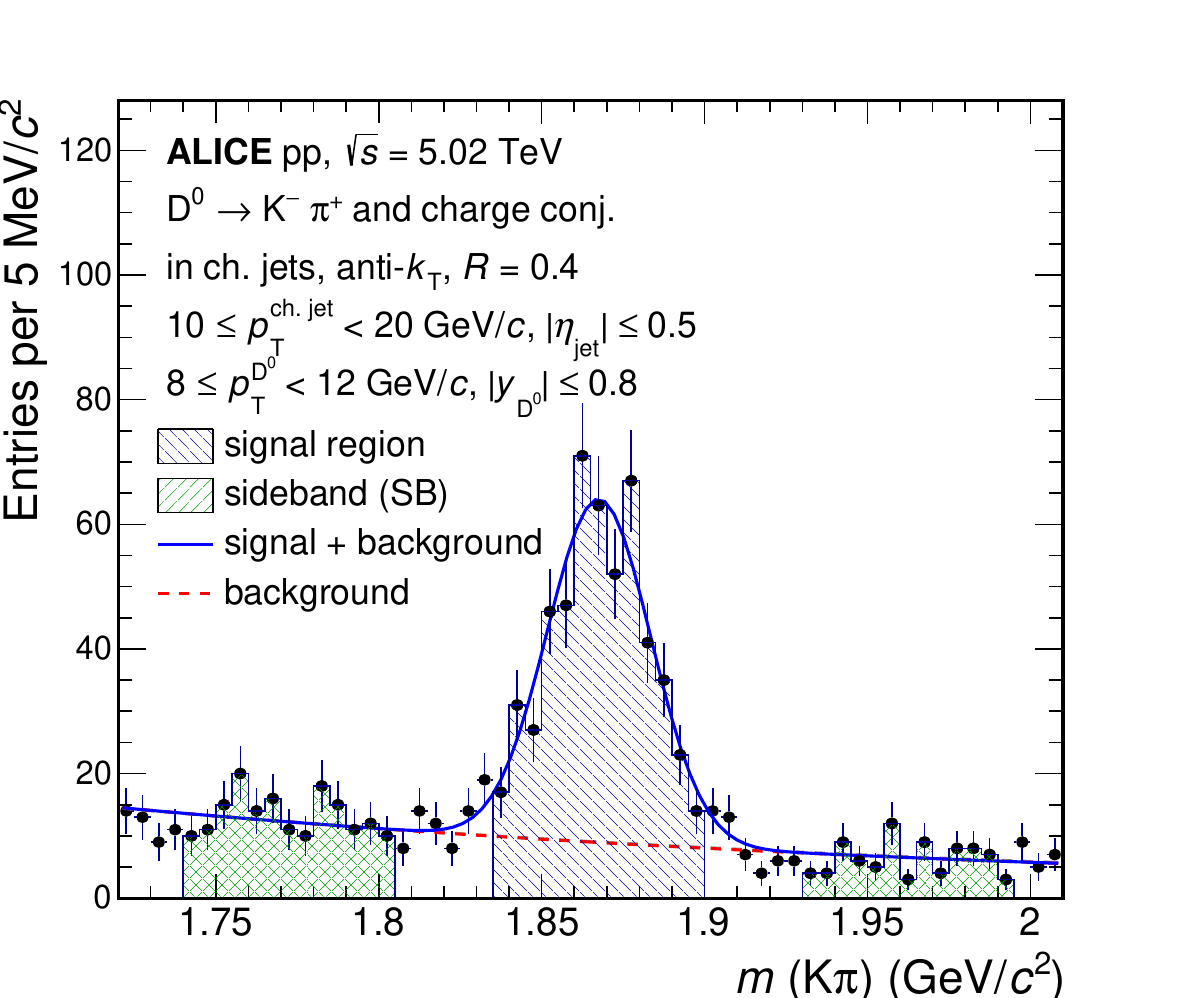}
\hspace{0cm}
\includegraphics[scale=0.37]{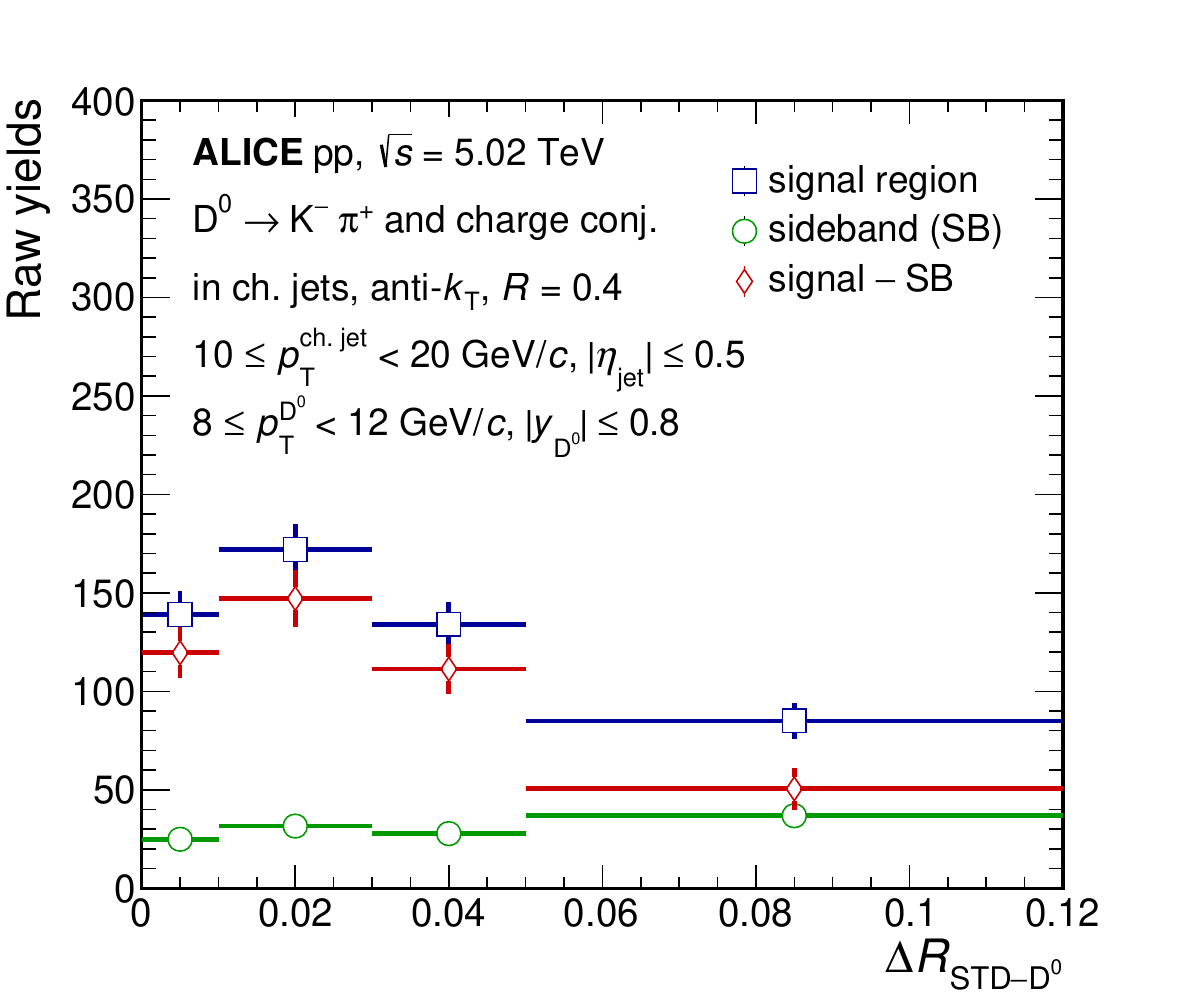}
\caption{Left: the \Dzero-decay candidate invariant-mass distribution for \jetptrange{10}{20} and $8 < p_{\rm T}^{\rm D^0} < 12$ \GeVc. The total fit function of the signal and background is represented by the blue line. The background fit function is represented by the red line. Right: the raw yields of the \Dzero-tagged jets as a function of \deltaR{STD}{\Dzero} in the signal region and sideband region.}
\label{fig:invmass}
\end{figure}

The invariant-mass distribution of the \Dzero candidates is extracted in intervals of \dzeropT and \jetpT. For each \dzeropT interval, these distributions are fitted with a Gaussian function in the signal region and an exponential function to define two sideband (background) shapes. See, as an example, a particular \jetpT and \dzeropT interval in Fig.~\ref{fig:invmass} (left). A fraction of signal ``reflection'' candidates exists, which pass the topological selections but can have a swapped mass assignment to the decay particles. This inconclusive PID results in the charged kaon and pion pair being accepted as both a \Dzero and a $\overline{\rm D^0}$ candidate. To remove these incorrect mass hypothesis assignments, MC-generated templates of reflection candidates are obtained and their contribution is added in the background fit.

After the contribution from reflections is removed, the sideband distributions are scaled to estimate and subtract the background in the signal region and extract the \Dzero signal. The signal region is constrained within $\pm 2 \sigma$ of the mean value of the Gaussian fit. The background regions are constrained to $4\sigma - 9 \sigma$ on either side of the Gaussian peak as this region, being far from the signal peak, is expected to be dominated by background candidates. The raw yields of the signal and background regions are shown in Fig.~\ref{fig:invmass} (right). Their subtracted yield is the measured jet axes difference spectra before corrections.

\subsection{Corrections}
\subsubsection{Reconstruction efficiency}
Once the combinatorial background is removed, what remains is a signal \Raxis distribution inside jets tagged with a \Dzero candidate. This must be corrected for the reconstruction efficiencies. The reconstruction efficiency is calculated from a charm-enriched MC sample anchored to the data sample, using the ratio of the \dzeropT spectra of the matched \Dzero-tagged jets at reconstruction level to the generated \Dzero-tagged jets. The truth-level (MC-generated) \Dzero mesons and their jets are required to be fully contained within the TPC acceptance and to have $\jetpT \geq 5$ GeV/$c$. The efficiency is calculated for prompt and non-prompt \Dzero mesons separately, and is strongly dependent on \dzeropT. This is because the \Dzero-meson topological selections are stricter at low \dzeropT so that the larger combinatorial background can be removed in that region. The raw yields, $N_{\rm raw}$, are corrected for the product of the acceptance and the prompt \Dzero (and $\overline{\rm D^0}$) jet reconstruction efficiency, $(\rm{Acc} \times \varepsilon)^{\rm c\rightarrow D^0}$, and then summed over all $p_{\rm T}^{\Dzero}$ intervals to obtain the total efficiency-corrected yield,

\begin{equation}
    N(\Rplain,\jetpT)= \sum_{\dzeropT} \frac{N_{\rm raw}(\Rplain,\jetpT,\dzeropT) }{({\rm Acc} \times \varepsilon)^{\rm c\rightarrow D^{0}} (\dzeropT)},
\end{equation}

where $N$ is the total efficiency-corrected yield and where \Rplain is the jet axes difference.

\subsubsection{Subtraction of non-prompt \Dzero mesons}\label{secsec:non-prompt}
At this point, the jet axes difference distributions still include non-prompt \Dzero mesons. These should be removed as they do not originate from a charm quark produced in the initial partonic interaction. No measurements of non-prompt \Dzero-tagged jets exist to date, so this contribution is estimated using NLO pQCD calculations from the POWHEG event generator~\cite{POWHEG_2010} coupled to the PYTHIA 8 generator~\cite{pythia_monash_2014} to describe the parton shower. The non-prompt 4-D response matrix, ${\rm RM}_{\rm det}^{\rm b \rightarrow D^{0}}(\detdeltaR,\truthdeltaR,\detjetpT,\truthjetpT)$, is applied to \textit{fold} the non-prompt distribution at truth level to detector level, which introduces detector effects in the distribution. The response matrix is estimated with the MC sample described in Sec.~\ref{sec:setupanddatasample}.

Before folding, the non-prompt \Dzero-tagged jet axes difference spectrum, $N_{\rm det}^{\rm b \rightarrow D^0}$, is corrected with the ratio of the non-prompt over prompt \Dzero efficiency (mimicking the same non-prompt \Dzero jet shape with prompt \Dzero jet efficiency in data), branching ratio and integrated luminosity, according to

\begin{equation}
    \begin{split}
    N_{\rm det}^{\rm b \rightarrow D^{0}}(\detdeltaR,\detjetpT) & = {\rm RM}^{\rm b \rightarrow D^{0}}_{\rm det}(\detdeltaR,\truthdeltaR,\detjetpT,\truthjetpT)\ 
    \\
     & \odot \sum_{\dzeropT} \frac{({\rm Acc} \times \varepsilon)^{\rm b\rightarrow D^0}(\dzeropT)}{({\rm Acc} \times \varepsilon)^{\rm c\rightarrow D^0}(\dzeropT)} N_{\rm POWHEG}^{\rm b \rightarrow D^{0}}(\truthdeltaR,\truthjetpT,\dzeropT ),
    \end{split}
    \label{eq:feeddowncorrection}
\end{equation}

where $N_{\rm POWHEG}^{\rm b \rightarrow D^0}$ is the number of jets with a \Dzero from beauty decay as estimated by POWHEG + PYTHIA~8, \detdeltaR and \truthdeltaR refer to the jet axes difference at detector and truth level, \detjetpT and \truthjetpT are the \jetpT at detector and truth level, and $(\rm{Acc} \times \varepsilon)^{\rm b\rightarrow D^0}$ and $(\rm{Acc} \times \varepsilon)^{\rm c\rightarrow D^0}$ signify the products of the acceptance and reconstruction efficiency for non-prompt-\Dzero and prompt-\Dzero jets, respectively. 

The folded non-prompt contribution is then subtracted from the efficiency-corrected jet axes difference distribution to obtain the corrected prompt distribution, $N_{\rm det}^{\rm c \rightarrow D^0}$, at detector level: 

\begin{equation}
    N_{\rm det}^{\rm c \rightarrow D^{0}}(\detdeltaR,\detjetpT) = N(\detdeltaR,\detjetpT) - N_{\rm det}^{\rm b \rightarrow D^{0}}(\detdeltaR,\detjetpT).
    \label{eq:feeddowncorrection}
\end{equation}

\begin{figure}[!htb]
\centering
\includegraphics[scale=0.5]{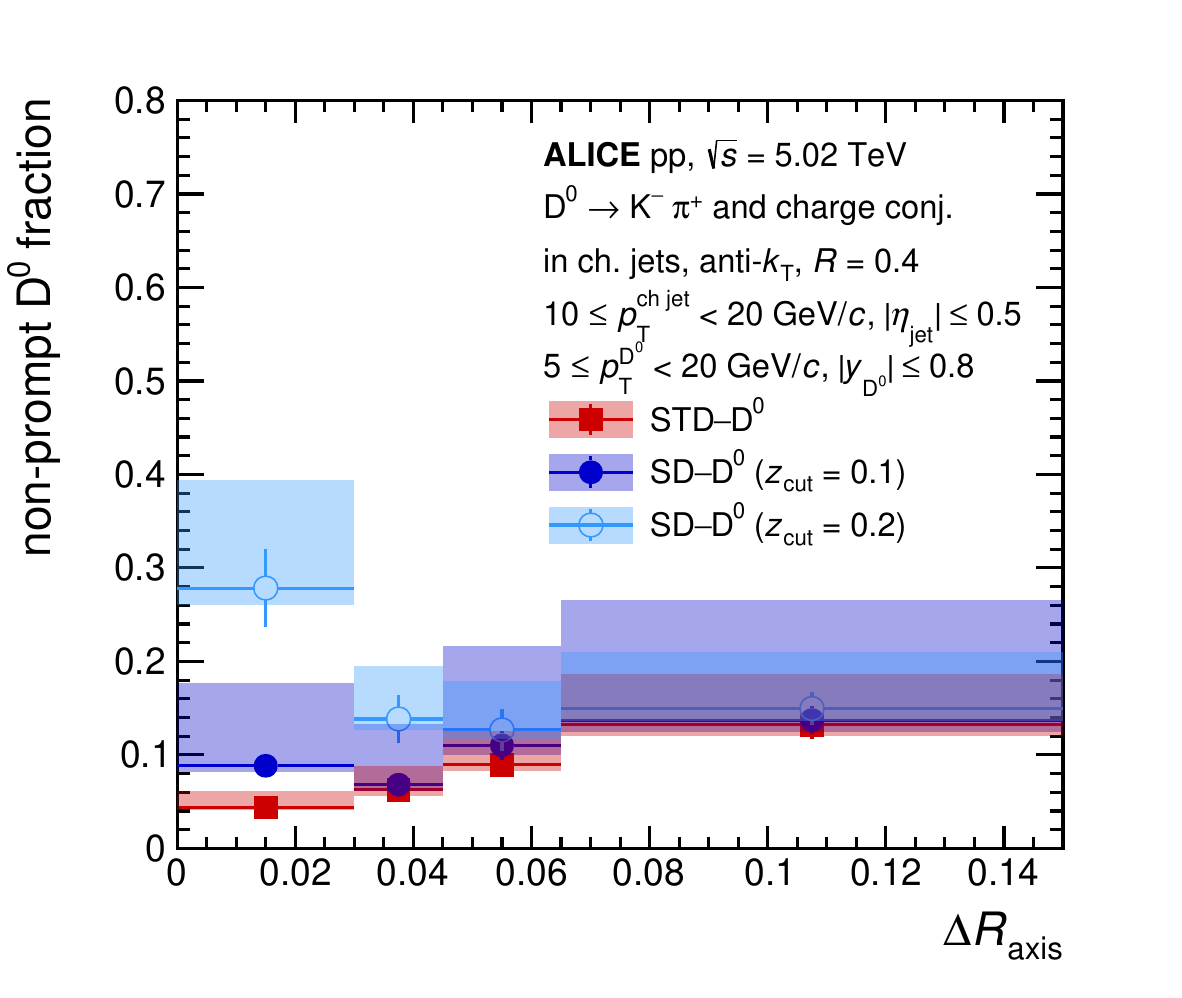}
\caption{Ratio of the simulated non-prompt \Dzero distribution in the detector (detector-level) over the efficiency-corrected data, which includes contributions from the prompt and feed-down \Dzero. The non-prompt \Dzero fraction is shown for \deltaR{STD}{\Dzero} and \deltaR{SD}{\Dzero} ($\zcut=0.1,0.2$ and $\beta=0$). Systematic and statistical uncertainties are represented by the corresponding color boxes and error bars, respectively.}
\label{fig:feeddownfraction}
\end{figure}

The ratios of the MC-generated non-prompt \Dzero mesons over the total prompt and non-prompt \Dzero mesons from data, are shown in Fig.~\ref{fig:feeddownfraction}, at detector level as a function of \Rplain. For jet axes difference distributions from the standard jet axis, this fraction is below 20$\%$. Non-prompt jets generally have a harder substructure that will pass a grooming procedure with $\zcut=0.2$. They have a harder splitting at a smaller angle, which can be explained by the additional decay products produced in the decay of the beauty hadron~\cite{dead_cone_2022}. Meanwhile, prompt \Dzero jets that occupy small \Rplain are less likely to survive Soft Drop (Sec.~\ref{secsec:groomed}) so, when the jets are groomed, the non-prompt fraction increases at small \Rplain.

\subsubsection{Unfolding procedure}
The final correction to the jet axes difference distributions is the correction for detector effects such as the track \pt resolution, tracking inefficiencies, and interactions between particles and the material in the detector volume. This was achieved using an iterative Bayesian approach~\cite{bayesianUnfolding}, implemented in the RooUnfold package~\cite{RooUnfold}. The 4-D prompt response matrix, ${\rm RM}^{\rm c \rightarrow D^{0}}(\detdeltaR,\truthdeltaR,\detjetpT,\truthjetpT)$, maps detector-level variables $(\detdeltaR,\detjetpT)$, reconstructed in the full PYTHIA 8 + GEANT 3 detector simulation (described in Sec.~\ref{secsec:non-prompt}), to the corresponding truth-level variables $(\truthdeltaR,\truthjetpT)$. The detector-level and truth-level jets are matched based on the requirement that they contain the same prompt \Dzero meson among their constituents. The RM is scaled by the prompt \Dzero-jet efficiency before unfolding the measured spectra to truth-level.

The \Raxis distributions are unfolded together with the \jetpT spectra in a 2-D unfolding method. Five unfolding iterations are needed to achieve convergence within 5\% between subsequent iterations while maintaining minimal statistical uncertainties. 

The stability of the unfolding procedure is evaluated with a series of closure tests, which are discussed in detail in Ref.~\cite{Cruz_Torres_2022}. The unfolding procedure yields good agreement between the compared distributions in each test. The binning configuration for each jet axes difference distribution is chosen to achieve acceptable unfolding convergence and good closure test performance. The final results shown in Sec.~\ref{sec:results} are fully unfolded.

\section{Systematic uncertainties}\label{sec:systematics}

Five sources of systematic uncertainty that affect the jet axes difference distributions are considered. For each source, the root-mean-square of all deviations is assigned as the systematic uncertainty, unless otherwise stated.

\begin {enumerate}

    \item {\bf $\rm \bf D^0$ topological selections}: the distributions of variables used in the \Dzero topological selections may be different between data and MC. To account for these discrepancies, the systematic uncertainty due to the chosen topological selections on the \Dzero meson are estimated using five variations of the selection criteria. These are obtained by varying the selections such that $\pm10\%$ and $\pm15\%$ variations in the selection efficiencies are induced.
    
    \item {\bf Yield extraction}: an uncertainty is assigned to the fitting strategy in the raw-yield extraction procedure, described in Sec.~\ref{subsec:yieldextraction}. The stability of this procedure is estimated by varying the signal extraction parameters and repeating the invariant-mass fits. The varied parameters include the assumed background fitting functions (exponential or linear), signal range, sideband extraction range, and the width and mean of the Gaussian signal.
    
    \item {\bf Tracking efficiency}: uncertainties on the track-reconstruction efficiency influence the jet-momentum resolution and \Dzero meson reconstruction efficiency. The uncertainty on the tracking efficiency is approximately 3\% in pp collisions, determined as a combination of uncertainties due to the TPC track-selection efficiency and the TPC--ITS matching efficiency. The effect of this uncertainty on the jet axes difference distributions is evaluated by randomly rejecting 3\% of the reconstructed tracks in the anchored MC and repeating the analysis with the new reconstruction efficiency and response matrices. 
    
    \item {\bf Non-prompt $\bf D^0$ contribution}: the non-prompt \Dzero contribution, determined in simulation, is varied to account for uncertainty in the modeling. The non-prompt contribution is recalculated with varied input parameters of the POWHEG + PYTHIA 8 simulation including the beauty-quark mass, factorization scale factor, renormalization scale factor, and the Parton Distribution Function (PDF) choice. The maximum deviation of these variations from the standard correction is taken as a systematic uncertainty.

    \item{\bf Unfolding}: the systematic uncertainty due to the unfolding procedure has four contributions.

    \begin{itemize}

        \item {\bf Variation of the regularization parameter, \textit{\textbf{n}}$_{\bf iter}$}: the unfolding is repeated with the number of iterations varied by $\pm2$ around the central value. The average difference with respect to the central value is taken as an uncertainty. 
    
        \item {\bf Choice of the prior}: the prior (the predicted, truth-level distribution) is scaled by $(p_\mathrm{T}^\mathrm{ch\;jet})^{\pm 0.5} \times (1 \pm 0.5 \times (2 \Rplain - 1))$. These variations are chosen because they substantially change the prior and demonstrate prior independence across a range that would be reasonably expected in differing calculations. The larger variation between the $\pm$ cases is used as the uncertainty due to the choice of prior.
    
        \item {\bf Variation of the truncation of the detector-level \textit{\textbf{p}}$_{\rm \bf T}^{\rm \bf ch~jet}$}: the lower $p_{\text{T, det}}^{\text{ch jet}}$ limit, originally at 5 \GeVc, is truncated by 1 \GeVc. The difference from the central value is taken as an uncertainty.
    
        \item {\bf Variation of the $\Delta$\textit{\textbf{R}$_{\rm axis}$} binning}: alternate binning schemes are constructed, with finer and coarser granularity than the main result, to account for uncertainties due to the choice of binning (of each \Rplain observable) during the unfolding procedure. The size of every bin is increased or decreased by 20\% of its original size. Variations that did not converge in unfolding quality tests are removed. The uncertainty is calculated by performing a linear fit of the ratio of each variation to the default value.
    
    \end{itemize}
    These four sources probe the same source of uncertainty, so the total unfolding uncertainty is estimated here as the standard deviation of all four contributions.
\end{enumerate}

The total systematic uncertainty is calculated by assuming the individual sources are uncorrelated uncertainties, and summing them in quadrature. A summary of the systematic uncertainties of all jet axes difference observables is provided in Tab.~\ref{tab:systematics}, but excludes the systematic uncertainties for \deltaR{WTA}{D^0}, where only the first bin is reported as detailed in Sec.~\ref{secsec:wta-d}.

\begin{table}[bth]
\small
\caption{Systematic uncertainties of the \Dzero-tagged jet axes difference measurements. The minimum and maximum values of the uncertainties across the \Rplain intervals are provided.
\label{tab:systematics}
}
\vspace{-0.8cm}
\begin{center}
\begin{tabular}{|m{7.5em}||cc|ccc|ccc|}
\hline \hline
\multicolumn{1}{c}{~} & \multicolumn{2}{c}{Standard Sample} & \multicolumn{3}{c}{Groomed ($\zcut=0.1$, $\beta=0$)} & \multicolumn{3}{c}{Groomed ($\zcut=0.2$, $\beta=0$)} \\
\hline
Systematic \mbox{Unc. Source} & STD--\Dzero & STD--WTA & SD--\Dzero & WTA--SD & STD--SD & SD--\Dzero & WTA--SD & STD--SD \\ 
\hline \hline
Tracking \mbox{Efficiency} & 0--5\% & 0--5\% & 1--5\% & 2--4\% & 1--14\% & 0--2\% & 0--1\% & 0--15\% \\ 
\hline
Non-Prompt \mbox{Variation} & 1\% & 1--2\% & 1--3\% & 1--3\% & 0--5\% & 2--7\% & 2--7\% & 2--10\% \\ 
\hline
Yield \mbox{Extraction~~~~~~} & 1--3\% & 1--3\% & 2--3\% & 2--4\% & 1--5\% & 3--9\% & 4--9\% & 2--11\% \\ 
\hline
\Dzero~\mbox{Topological~~~~~}
\mbox{Selections} & 1--3\% & 2--4\% & 1--4\% & 1--4\% & 4--22\% & 3--14\% & 3--12\% & 4--17\% \\ 
\hline
Unfolding \mbox{Variations} & 1\% & 2\% & 2\% & 1\% & 9\% & 5\% & 5\% & 11\% \\ 
\hline
\hline

Total~\mbox{Systematic~~~} \mbox{Uncertainty~~~~~} & 3--6\% & 3--7\% & 4--6\% & 4--7\% & 10--28\% & 9--16\% & 8--14\% & 12--29\% \\ 
\hline \hline
\end{tabular}
\end{center}
\end{table}

\section{Results and discussion}\label{sec:results}

The fully corrected and differential jet axes difference distributions for \Dzero-tagged jets are reported for $R=0.4$ anti-$k_{\rm T}$ jets. The distributions are normalized with respect to the total number of Standard, ungroomed jets. With the exception of the results reported in Tab. 2, this paper focuses on jets within \jetptrange{10}{20}. 

\subsection{Difference between the Winner-Takes-All and $\rm \bf D^{0}$ axes}\label{secsec:wta-d}
\begin{figure}[!ht]
\hspace{-1cm}
\hspace{0.1cm}
\centering
\includegraphics[width=0.8\linewidth]{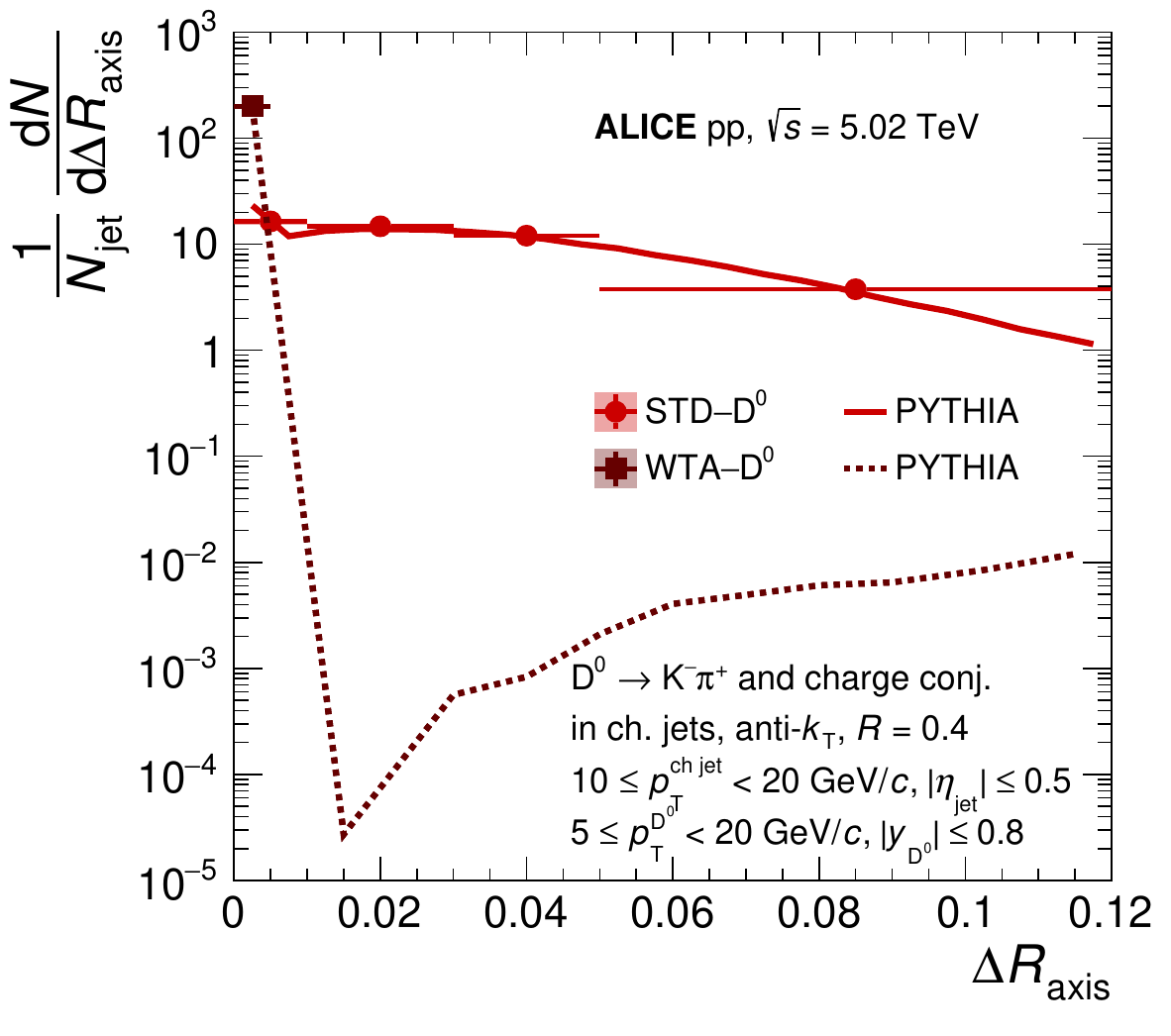}
\caption[]{Unfolded jet axes difference distribution for \deltaR{WTA}{\Dzero} and \deltaR{STD}{\Dzero} in \jetptrange{10}{20}. Includes systematic and statistical uncertainties represented by color boxes and error bars, respectively, and a comparison to predictions from PYTHIA 8.}	
\label{fig:unf_wta_d}
\end{figure}

The difference between the Winner-Takes-All and \Dzero axes, \deltaR{WTA}{\Dzero}, is shown in Fig.~\ref{fig:unf_wta_d}, together with the result for \deltaR{STD}{\Dzero}, which will be discussed in more detail in the next section. While the \deltaR{STD}{\Dzero} distribution peaks at low \Rplain, the Standard jet axis may not always align with the \Dzero~meson. Conversely, \deltaR{WTA}{\Dzero} peaks strongly at $\Rplain=0$. In fact, only the first interval is reported because the number of counts outside this interval is too small to unfold robustly. The reported interval matches the PYTHIA 8 prediction, which also shows very few counts outside the smallest \Rplain interval. 

Table~\ref{tab:wta_dzero} presents the total fraction of \Dzero-tagged jets in $0<\Rplain<0.005$, over the total number of \Dzero-tagged jets found within $R=0.4$. The central value of the measurement is 99\% for the lower \jetpT interval and 95\% for the higher \jetpT interval, with systematic uncertainties of 1\% and 5\%, respectively. Across both \jetpT regions, the measurements are consistent within uncertainties.

%%%%%%%%%%%%%%%%%%%%%%%%%%%%%%%

\begin{table}[bth]
\caption{Fraction of \Dzero-tagged jets where the difference between the Winner-Takes-All jet axis and \Dzero is less than 0.005. ``Measurement'' reports the observed central value with the corresponding statistical uncertainties. A statistical uncertainty $< 1\%$ is considered negligible. ``Sys. Uncertainty'' reports the systematic uncertainties in each \jetpT interval, and ``PYTHIA 8" reports the same fraction modeled in PYTHIA 8 with the corresponding statistical uncertainties.}
\label{tab:wta_dzero}
\begin{center}
\begin{tabular}{lrrrrr}
\multicolumn{3}{c} {Fraction of jets in $0 \leq \Rplain \rm < 0.005$ for \deltaR{WTA}{D^0}} \\
\hline
Distribution & \multicolumn{1}{c} {\jetptrange{10}{20}} & \multicolumn{1}{c} {\jetptrange{20}{50}} \\ 
 ~ & $5 \leq \textit{p}_{\rm T}^{\Dzero} < 20~\GeVc$ & $12 \leq \textit{p}_{\rm T}^{ \Dzero} < 36~\GeVc$ \\ \hline
Measurement & $99\%~({\rm negl.~stat.~unc.})$ & $(95 \pm 2) \%$ \\ 
Sys. Uncertainty & $\hspace{0.1cm} \pm 1\%$ & $\hspace{0.1cm} \pm 5\%$   \\ 
PYTHIA~8 & $99.5\%~({\rm negl.~stat.~unc.})$ & $99.8\%~({\rm negl.~stat.~err.})$  \\
\hline
\end{tabular}
\end{center}
\end{table}
%%%%%%%%%%%%%%%%%%%%%%%%%%%%%%%%%%%%

It is, however, clear that the Winner-Takes-All and the \Dzero axes align extremely well. Since the Winner-Takes-All jet algorithm tracks the hardest particle in the hardest branch of the jet, it is insensitive to soft radiation and recoil effects~\cite{wta_Larkoski}. Thus, this result demonstrates that the \Dzero meson is the leading particle in the jet. This result is expected in heavy-flavor jets due to the presence of the dead cone, which allows the \Dzero meson to retain a large fraction of the total jet momentum and increases its likelihood of being selected as the leading particle. Previous studies on the \Dzero-tagged momentum fraction, $z_{||}$, showed that the \Dzero is very often the leading particle ($z_{||} > 0.5$)~\cite{d0_prod_5and13TeV_2023}. However, these studies are constrained by the reported $z_{||}$ range and do not explore the regions where the \Dzero may be a sub-leading particle. While \deltaR{WTA}{\Dzero} is also bounded in $z_{||}$, given $p_{\rm T}^{\Dzero} > 5$ \GeVc, a subset of the measured jets in the sample may access down to $z_{||} = 0.25$, where the \Dzero is not necessarily leading. Consequently, the fraction reported in Tab.~\ref{tab:wta_dzero} quantifies the probability that the \Dzero is leading for $R=0.4$ jets with \jetptrange{10}{50} and $5 \leq \textit{p}_{\rm T}^{\Dzero} < 36~\GeVc$.

\subsection{Comparison of $\rm \bf D^{0}$, Standard, and Winner-Takes-All jet axes}
\label{secsec:std-d}

\begin{figure}[!ht]
\hspace{-1cm}
\centering
\includegraphics[width=1.05\linewidth]{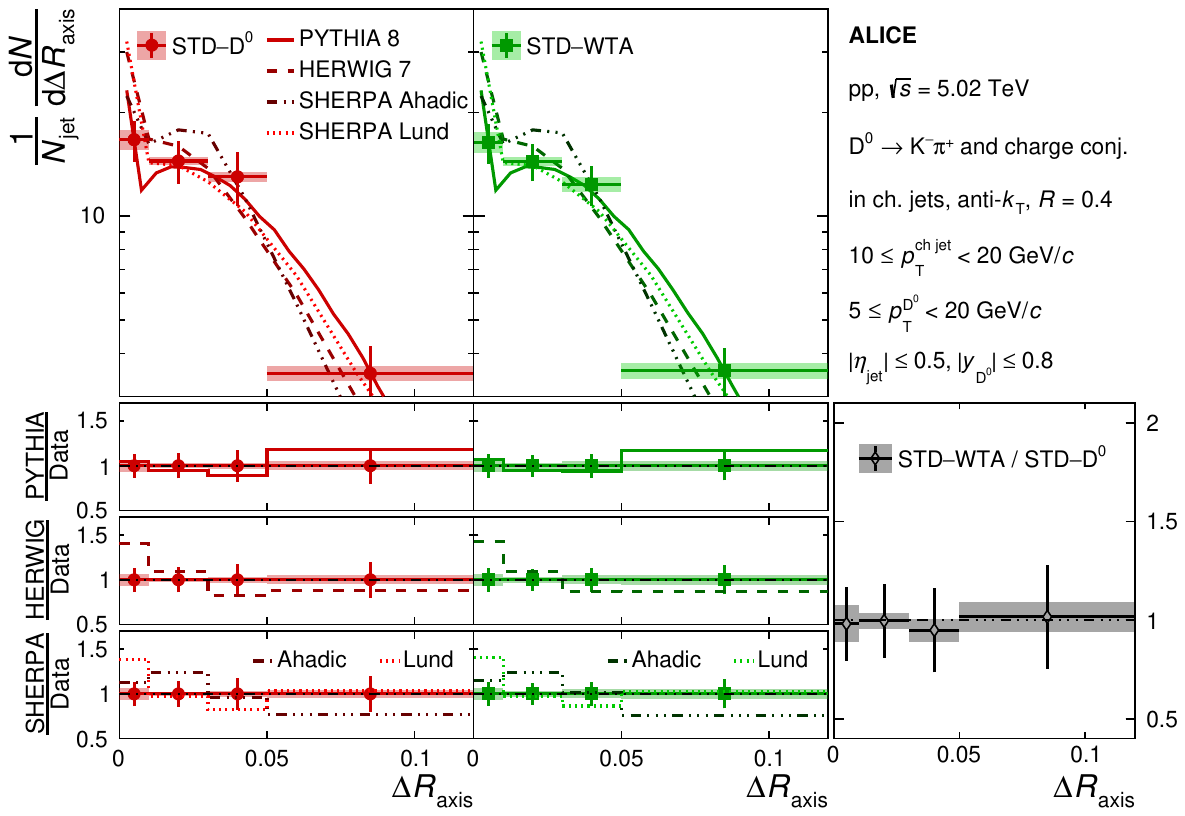}
\caption[]{Fully unfolded jet axes difference distribution for \deltaR{STD}{\Dzero} (left) and \deltaR{STD}{WTA} (middle) for \jetptrange{10}{20}. Systematic and statistical uncertainties are represented by color boxes and error bars, respectively, and comparisons to MC event generators PYTHIA 8, HERWIG 7, and SHERPA 2 (Ahadic and Lund) are shown in the respective bottom panels. The bottom right panel shows a ratio of the two data distributions.}	
\label{fig:unf_std-d+std-wta}
\end{figure}
%%%%%%%%%%%%%%%%%%%%%%%%%%%%%%%

The difference between the Standard and \Dzero axes, \deltaR{STD}{\Dzero}, and the difference between the Standard and the Winner-Takes-All axes, \deltaR{STD}{WTA}, are shown in Fig.~\ref{fig:unf_std-d+std-wta} for \jetptrange{10}{20}. The normalized distributions for \deltaR{STD}{D^0} and \deltaR{STD}{WTA} match extremely well due to the observed alignment between the Winner-Takes-All axis and the \Dzero direction. Their ratio is consistent with unity within uncertainties, with a $\chi ^2$ test yielding a p-value of 0.99, indicating a strong agreement.

Model comparisons show that PYTHIA 8 reproduces the observed data within experimental uncertainties. The sharp spike at small \Rplain, visible with the fine granularity of the model, is discussed in Sec.~\ref{secsec:groomed}. SHERPA Lund~\cite{sherpa2}, utilizing a string-based hadronization similar to PYTHIA, aligns with the data outside the smallest \Rplain region, where it slightly diverges. In contrast, the shapes of HERWIG~\cite{herwig7} and SHERPA Ahadic~\cite{sherpa2}, which rely on cluster-based hadronization, are steeper than the data over the full range of \Rplain.

\subsection{Groomed jet axes differences}
\label{secsec:groomed}

\begin{figure}[!ht]
\hspace{-1cm}
\hspace{0.1cm}
\vspace{-0.5cm}
\centering
\includegraphics[width=0.8\linewidth]{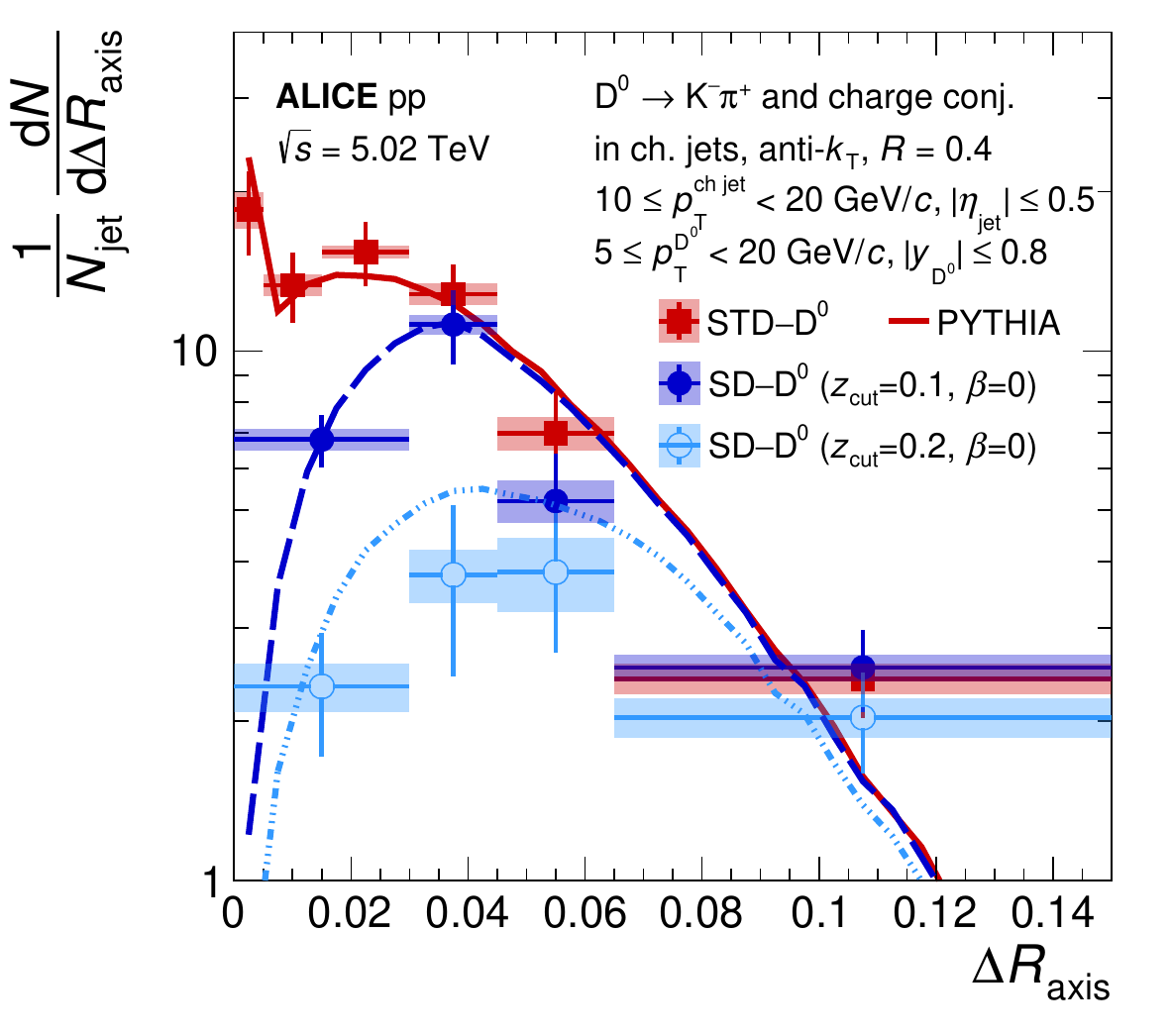}
\caption[]{Fully unfolded jet axes difference distributions for \deltaR{STD}{\Dzero} and \deltaR{SD}{\Dzero}, where the Soft Drop parameters are $\beta=0$ and $\zcut = 0.1$ or $0.2$, in \jetptrange{10}{20}. Systematic and statistical uncertainties are represented by color boxes and error bars, respectively. Comparisons to predictions from PYTHIA 8 are included.}	
\label{fig:unf_STDandSD}
\end{figure}
%%%%%%%%%%%%%%%%%%%%%%%%%%%%%%%

\begin{figure}[!ht]
\hspace{-1cm}
\centering
\includegraphics[width=1.05\linewidth]{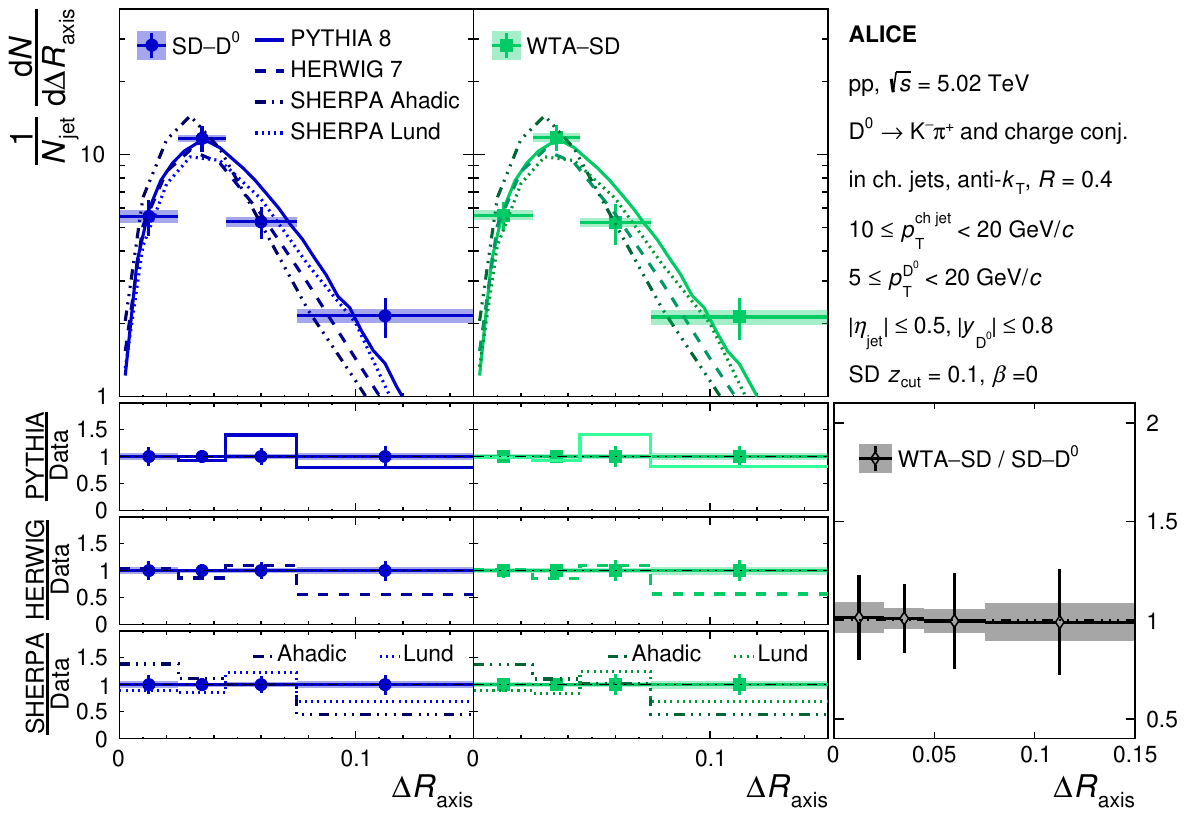}
\caption[]{Fully unfolded jet axes difference distributions for \deltaR{SD}{\Dzero} (left) and \deltaR{WTA}{SD} (middle), with grooming parameters $\zcut=0.1$ and $\beta=0$, for \jetptrange{10}{20}. Systematic and statistical uncertainties are represented by color boxes and error bars, respectively, and MC event generator comparisons are shown in the bottom panels. The bottom right panel shows a ratio of the two data distributions.}	
\label{fig:unf_sd-d+wta-sd_z01}
\end{figure}
%%%%%%%%%%%%%%%%%%%%%%%%%%%%%%%

\begin{figure}[!ht]
\hspace{-1cm}
\centering
\includegraphics[width=1.05\linewidth]{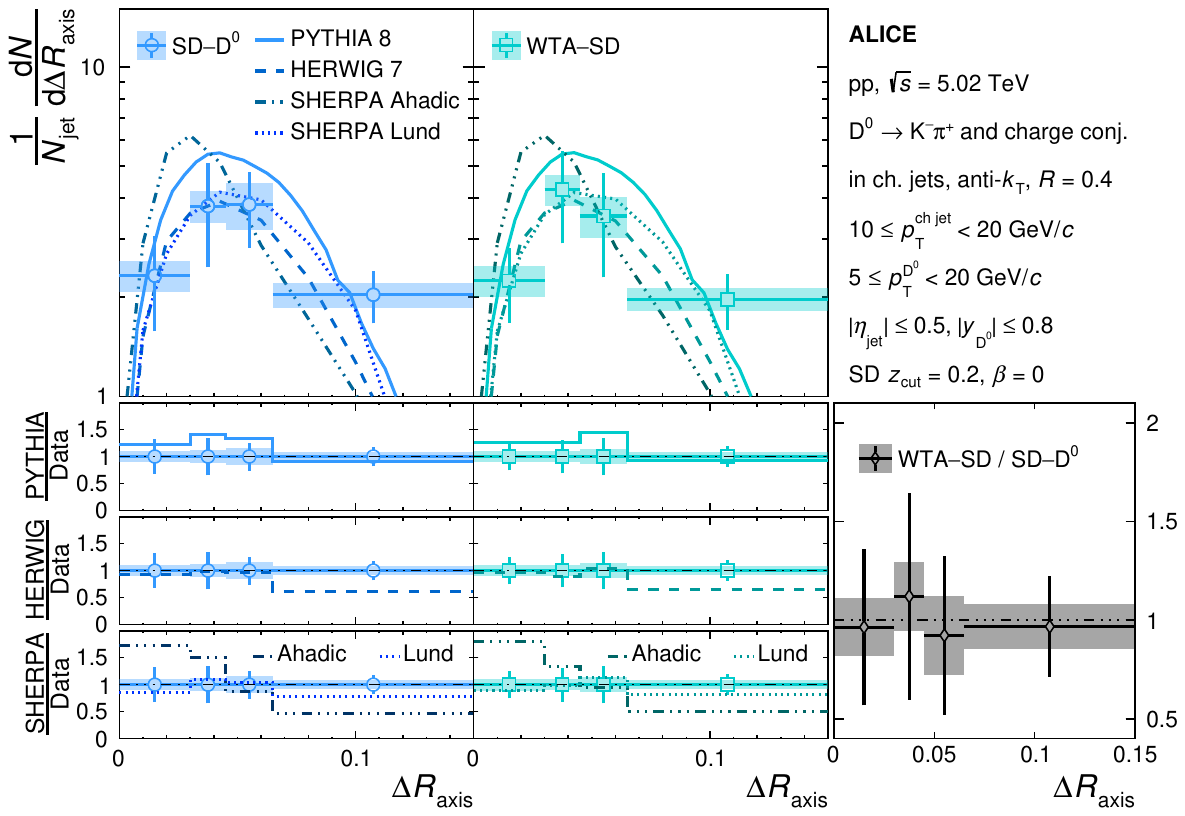}
\caption[]{Fully unfolded jet axes difference distributions for \deltaR{SD}{\Dzero} (left) and \deltaR{WTA}{SD} (middle), with grooming parameters $\zcut=0.2$ and $\beta=0$, for \jetptrange{10}{20}. Systematic and statistical uncertainties are represented by color boxes and error bars, respectively, and MC event generator comparisons are shown in the bottom panels. The bottom right panel shows a ratio of the two data distributions.}	
\label{fig:unf_sd-d+wta-sd_z02}
\end{figure}
%%%%%%%%%%%%%%%%%%%%%%%%%%%%%%%

Grooming jets with the Soft Drop algorithm removes the softest radiation, making the jet properties less dependent on the non-perturbative part of the jet evolution. In this work, the angular weight parameter $\beta$ is held at zero, causing the Soft Drop groomer to remove all radiation softer than the $z_{\rm cut}$ value.

To illustrate the impact of Soft Drop on jet axes differences, Fig.~\ref{fig:unf_STDandSD} compares the Standard and groomed jet samples. In this figure, \deltaR{STD}{\Dzero} is extended to a finer granularity in the smallest \Rplain interval. The sharp spike near $\Delta R = 0$, which deviates from the general shape of \deltaR{STD}{\Dzero}, contains jets that either have high momentum asymmetry or a single charm quark that does not radiate at all. These jets have few or no splittings, and are more likely to be groomed away when the Soft Drop condition is applied. This is confirmed by the distribution of \deltaR{SD}{\Dzero}. Fewer jets pass Soft Drop at small \Rplain, and fewer still when the grooming condition is stricter ($z_{\rm cut}=0.2$). This is a direct consequence of the dead-cone effect. Just as the charm meson maintains a large fraction of the jet momentum, any radiation emitted close to it is likely to be extremely soft, making the jet less likely to satisfy the grooming criteria at small \Rplain.

As \Rplain increases, the \deltaR{SD}{\Dzero} distributions, first $z_{\rm cut}=0.1$ and then $z_{\rm cut}=0.2$, begin to agree until all three curves are statistically indistinguishable. This larger \Rplain region is populated by more symmetric jets (i.e. the prong structure of the jets is more similar in momentum) that are not as sensitive to the Soft Drop condition and are more likely to survive grooming.

Figures~\ref{fig:unf_sd-d+wta-sd_z01} and~\ref{fig:unf_sd-d+wta-sd_z02} show
\deltaR{SD}{\Dzero} and \deltaR{WTA}{SD} for \jetptrange{10}{20}, to explore the effect of grooming on jet axes differences. The strongest grooming, with $\zcut=0.2$, is shown in Fig.~\ref{fig:unf_sd-d+wta-sd_z02}, while Fig.~\ref{fig:unf_sd-d+wta-sd_z01} illustrates less intense grooming, with $\zcut=0.1$.
The ratio between \deltaR{SD}{\Dzero} and \deltaR{WTA}{SD} is also reported for both cases of $\zcut$. The ratio is consistent with unity for both cases of $\zcut$, with a p-value of 0.99. This p-value holds for both grooming settings, as expected from the observed alignment between the Winner-Takes-All axis and the \Dzero. 

The jet axes differences are fairly well reproduced by PYTHIA 8 and SHERPA Lund for  $\zcut=0.2$, but SHERPA Lund and, considering the statistical errors, PYTHIA do not do as well for $\zcut=0.1$, suggesting that the remaining soft splittings are challenging for the models which employ string-breaking descriptions of hadronization. HERWIG reproduces the data somewhat better for both values of $\zcut$. SHERPA Ahadic does not describe the data. The performance of HERWIG 7 and SHERPA Ahadic does not change with increased grooming $\zcut$. Overall, the two string-based models, PYTHIA and SHERPA Lund, do agree better with the data. Although HERWIG is cluster-based, it has a different parton-shower prescription than SHERPA Ahadic, and describes groomed jets better.

\begin{figure}[!ht]
\hspace{-1cm}
\hspace{0.1cm}
\centering
\includegraphics[width=0.8\linewidth]{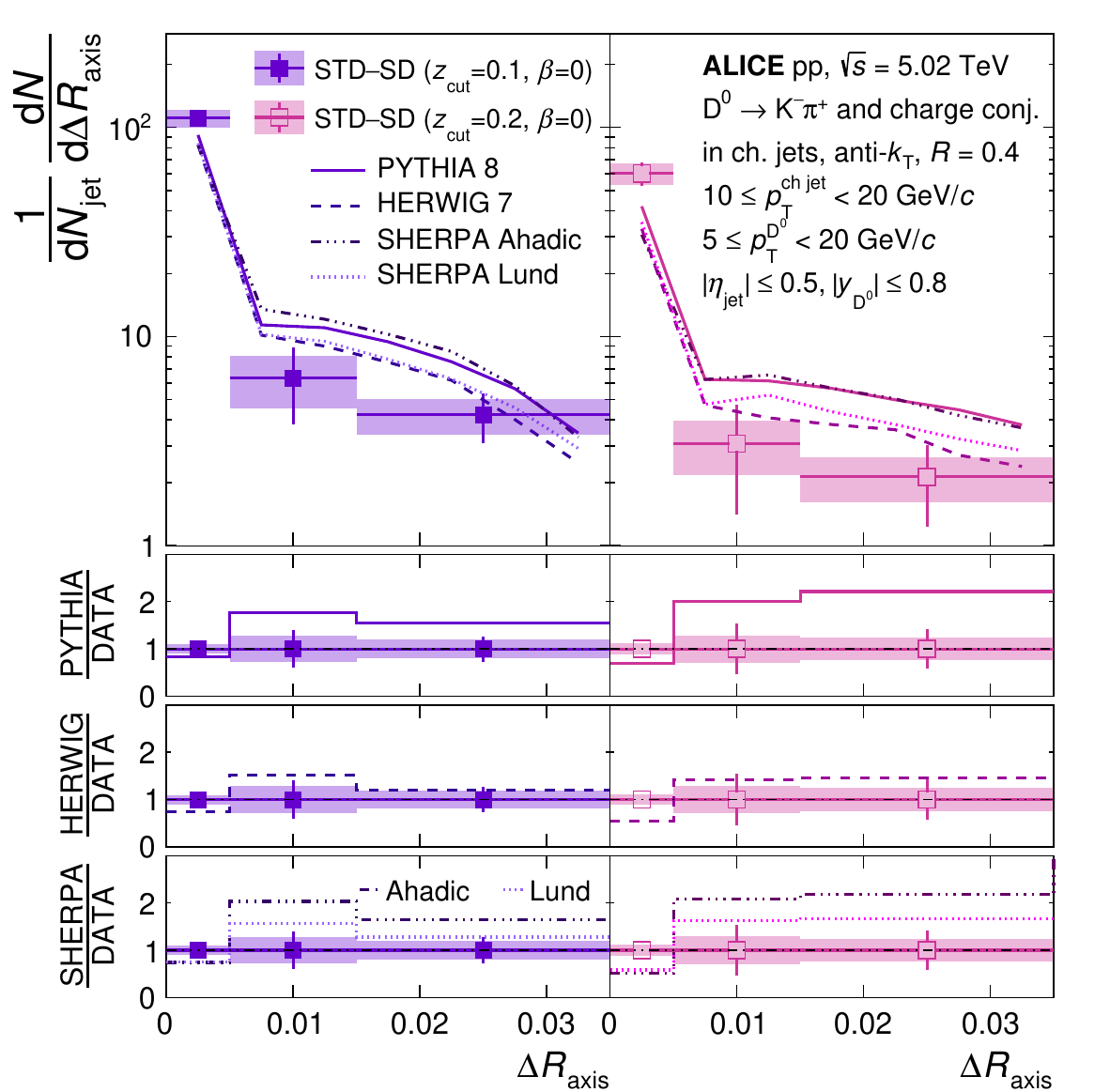}
\caption[]{Fully unfolded jet axes difference distributions for \deltaR{STD}{SD} with $z_{\rm cut}=0.1$ (left) and $z_{\rm cut}=0.2$ (right), $\beta=0$ and \jetptrange{10}{20}. Systematic and statistical uncertainties are represented by color boxes and error bars, respectively, and MC event generator comparisons are shown in the bottom panels.}	
\label{fig:unf_std_sd}
\end{figure}

Figure~\ref{fig:unf_std_sd} shows \deltaR{STD}{SD} with \jetptrange{10}{20} for both values of $\zcut$. The distributions peak sharply in the first \Rplain interval. For jets outside this interval, grooming away the soft radiation shifts the direction of the jet axis away from the Standard axis.

Interestingly, the models generally provide a better description of \deltaR{STD}{SD} when $\zcut=0.1$. This outcome is unexpected, as stronger grooming (higher \zcut) enhances the dominance of jets containing more perturbative radiation, which typically improves a model's description of the data. However, for all cases of \zcut, HERWIG and SHERPA Lund offer the best agreement with the data. For slightly higher \jetpT, HERWIG also reproduces the inclusive \deltaR{STD}{SD} distribution quite well~\cite{Cruz_Torres_2022}.

%%%%%%%%%%%%%%%%%%%%%%%%%%%%%%%

For \jetptrange{20}{40} inclusive jets without a \Dzero tag, \deltaR{WTA}{SD} shows that grooming did not substantially change the jet direction~\cite{Cruz_Torres_2022}. In \Dzero-tagged jets, the presence of the dead cone suppresses small-angle splittings such that the \Dzero takes a larger fraction of the jet's total momentum than the ``winner" in inclusive jets. This results in stronger sensitivity to the Soft Drop condition as a larger fraction of \Dzero-tagged jets do not satisfy Soft Drop compared to the inclusive case. The axis shifts more than in the inclusive case because the \Dzero~axis always remains inside the core of the jet, regardless of the grooming. Consequently, it is always a hard, wide-angle splitting from the \Dzero that is drawing the Soft Drop axis away from the jet core.

\begin{figure}[!ht]
\centering
\vspace{-1.7cm}
\includegraphics[trim={0 3.2cm 0 0},clip,width=0.79\linewidth]{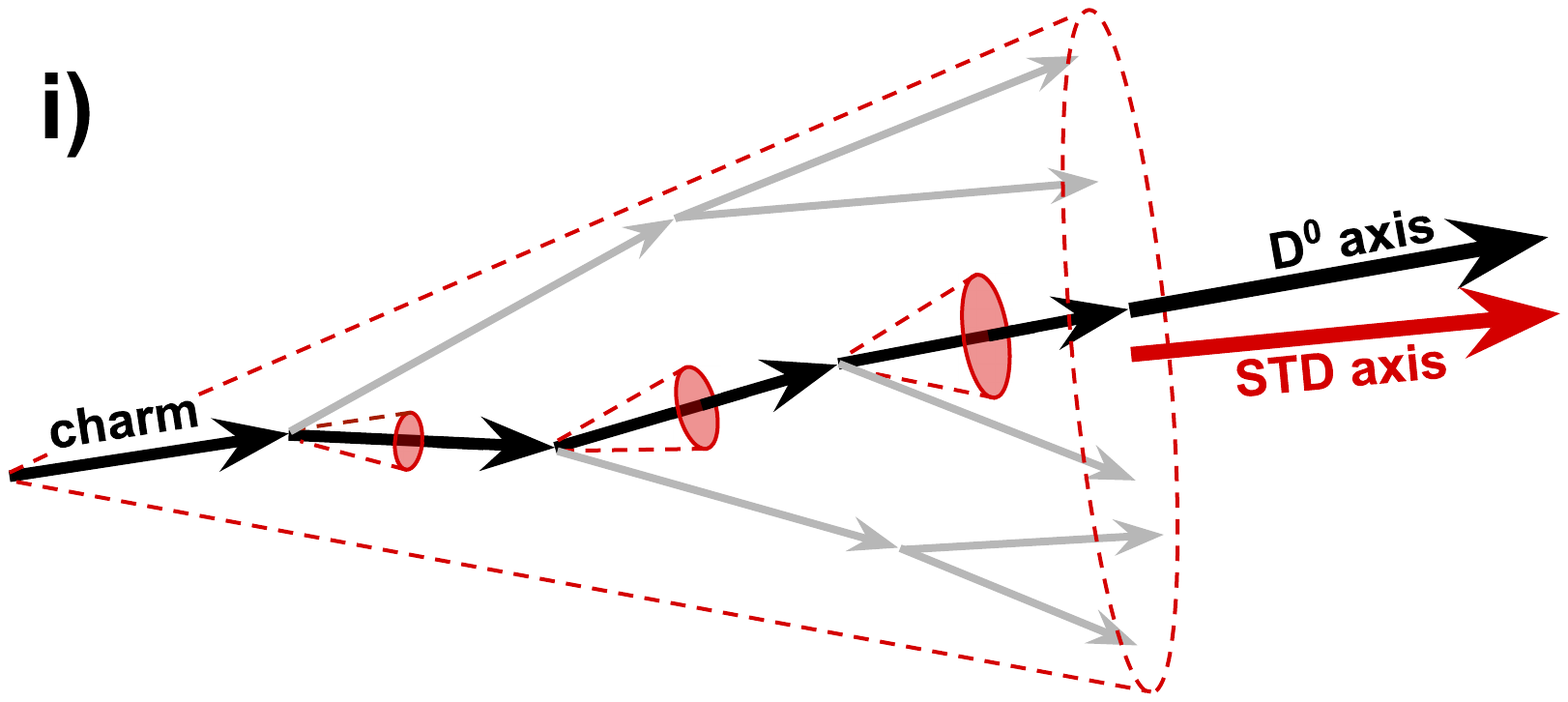}

\vspace{-2.5cm}

\includegraphics[trim={0 3cm 0 0},clip,width=0.79\linewidth]{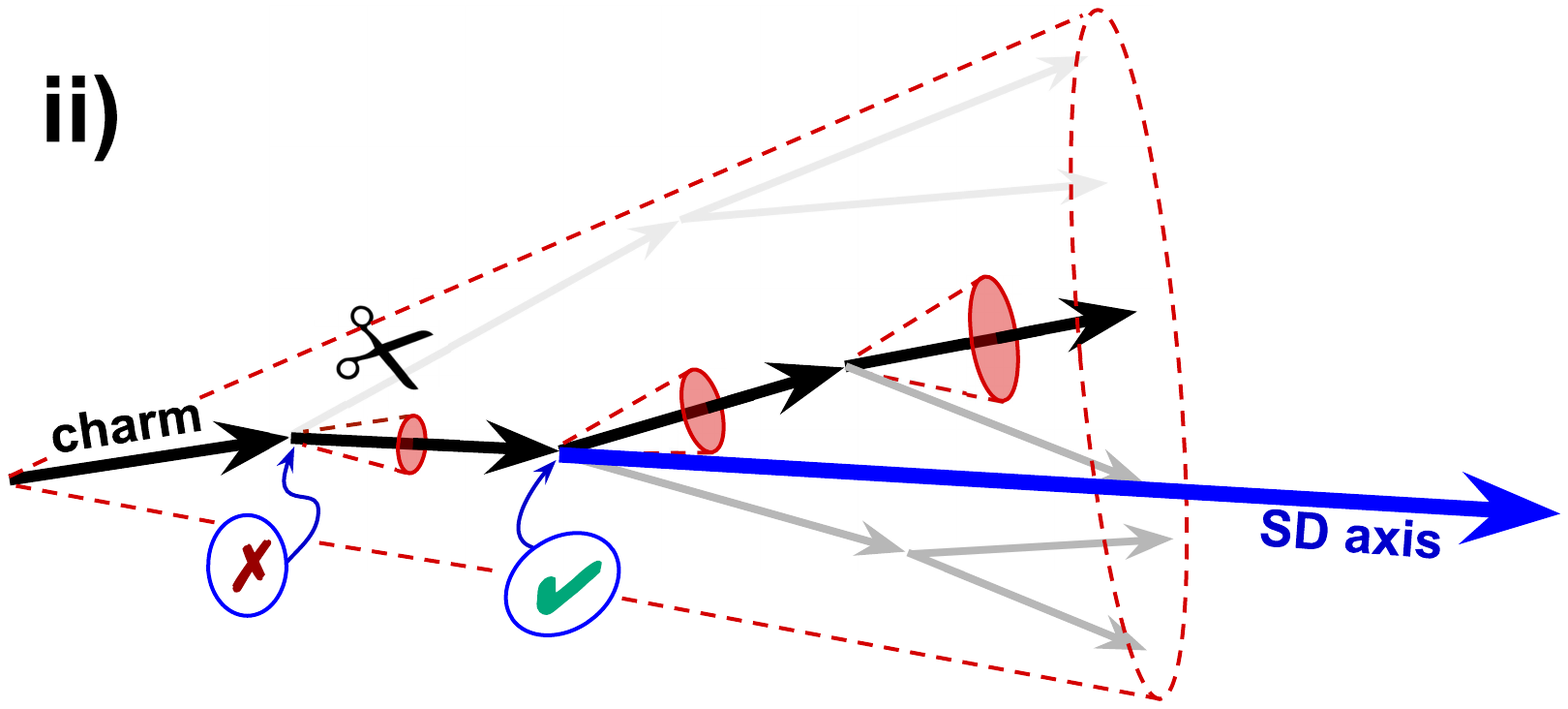}

\vspace{-2.5cm}

\includegraphics[trim={0 4cm 0 0},clip,width=0.79\linewidth]{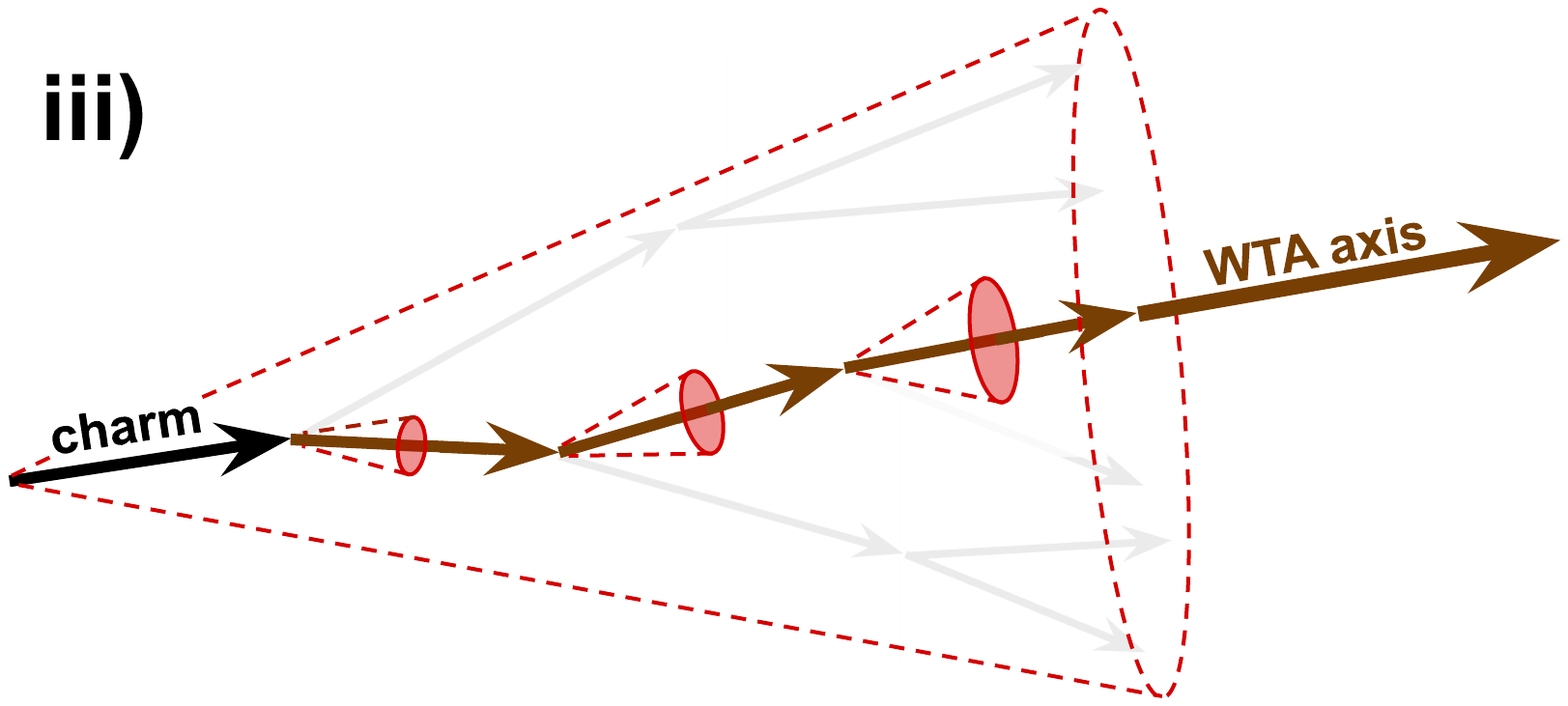}

\vspace{-0.7cm}

\caption[]{A representation of a prompt \Dzero-tagged jet, with the three jet axes (Standard, Soft Drop, and Winner-Takes-All) and the \Dzero axis. The dead cone is represented by the red cones around the charm quark at each splitting. i) The Standard jet axis is calculated as the sum of the jet constituent four-momenta. ii) The Soft Drop jet axis is determined from the first splitting which passes the Soft Drop condition. In this example, it is satisfied by the second splitting following the charm quark. The first splitting does not satisfy the Soft Drop condition, thus the softer branch is removed. iii) The Winner-Takes-All jet axis aligns with the hardest subjet in the hardest branch at each splitting.}	
\label{fig:pictorial}
\end{figure}

To construct a clear summary of these results, Figure~\ref{fig:pictorial} illustrates how each of the chosen jet axes are situated in an average prompt \Dzero-tagged jet. The charm quark initiates the jet, and the dead cone, which grows with each splitting as the charm quark loses energy, suppresses collinear radiation. The Standard axis results from recombining the constituents of the jet using the $E$-scheme (Sec.~\ref{sec:introduction}). As found in Sec.~\ref{secsec:std-d}, it is located at a small angle from the \Dzero meson. The Soft Drop axis is calculated by checking each splitting along the harder branch against the Soft Drop condition. It is less sensitive to soft radiation than the Standard axis and is often the furthest from the jet core. Lastly, the Winner-Takes-All axis is insensitive to soft radiation and, as discussed in Sec.~\ref{secsec:wta-d}, is nearly always aligned with the \Dzero meson.

%%%%%% Pictorial of STD, WTA and SD axes in the jet %%%%%%%%%%%
%%%%%%%%%%%%%%%%%%%%%%%%%%%%%%

\subsubsection{Probing flavor effects with PYTHIA 8}

\begin{figure}[!ht]
\hspace{-1.7cm}
\centering
\includegraphics[trim={0 1.5cm 0 0},clip,width=1.1\linewidth]{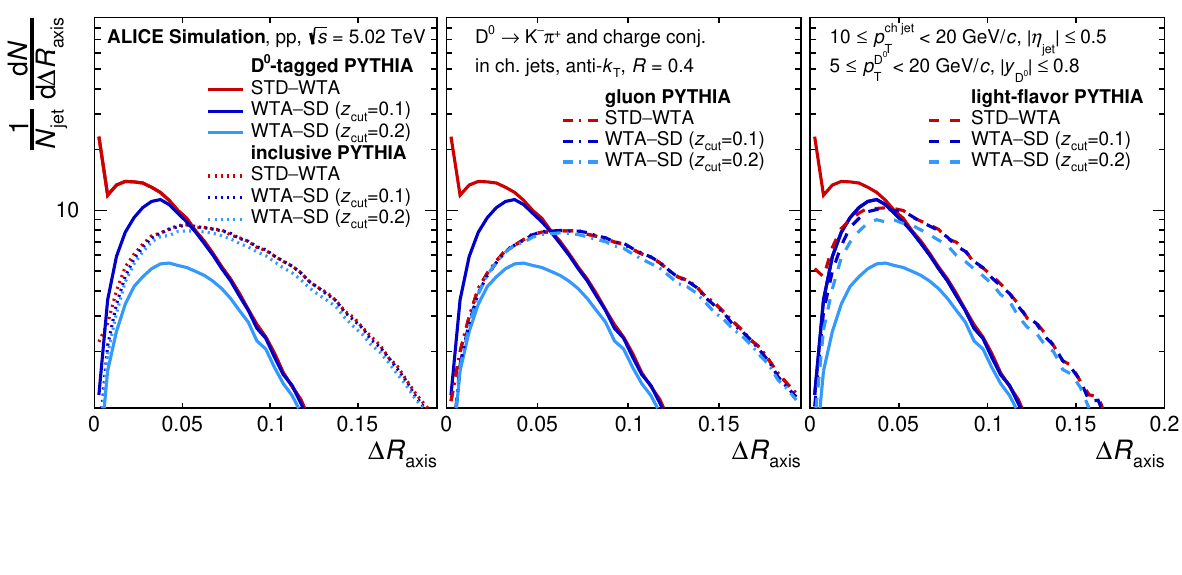}
\caption[]{PYTHIA 8 predictions of the jet axes difference distributions for \deltaR{STD}{WTA} and \deltaR{WTA}{SD}, where the Soft Drop parameters are $\beta=0$ and $\zcut = 0.1$ or $0.2$, in \jetptrange{10}{20}. The top panels show the normalized distributions. 
Left: \Dzero-tagged jets compared to inclusive jets. Middle: \Dzero-tagged jets compared to gluon-initiated jets. Right: \Dzero-tagged jets compared to light-quark-initiated jets. }	
\label{fig:STDandSD_gil}
\end{figure}
%%%%%%%%%%%%%%%%%%%%%%%%%%%%%%%

Predictions from PYTHIA 8 of jet axes differences, \deltaR{STD}{WTA} and \deltaR{WTA}{SD}, of various jet types are shown in Fig.~\ref{fig:STDandSD_gil}. The left-most panel shows a direct comparison of \Dzero-tagged jets to the inclusive (gluon-dominated) sample of jets in \jetptrange{10}{20}, probing both mass and Casimir-color effects. Casimir-color effects physically broaden the shape of the inclusive \Rplain distributions. Additionally, we observe that \Dzero-tagged jets are less likely to survive Soft Drop grooming than inclusive jets, as the dead cone prevents harder collinear radiation from forming at small \Rplain. 

The middle panel illustrates the difference between \Dzero-tagged jets (a quark-enriched sample) and gluon-initiated jets, where Casimir-color effects are examined. The \Rplain distribution from gluon-initiated jets has a slightly broader range than the inclusive-jet result, and is considerably broader than the light-quark initiated result. The right-most panel shows the difference between \Dzero-tagged jets and jets initiated by a light quark, which reveals the effect of the quark mass on the distribution. The light-flavor predictions show an increased sensitivity to the Soft Drop condition yet still less than the charm-jet predictions, which is expected given the mass of the charm quark relative to light quarks. A second peak is observed in the smallest \Rplain region of \deltaR{STD}{WTA} for both charm jets and light-flavor jets. This peak is due almost entirely to single-track jets, where the quark initiating the shower does not radiate at all. This peak is more pronounced in charm jets, suggesting that it is sensitive to the dead-cone effect and that future studies of single-track jets may provide a new way of estimating this prominent effect.

\subsubsection{Tuning the sensitivity to soft radiation with Soft Drop}

\begin{figure}[!ht]
\hspace{-1cm}
\hspace{0.1cm}
\centering
\includegraphics[width=\linewidth]{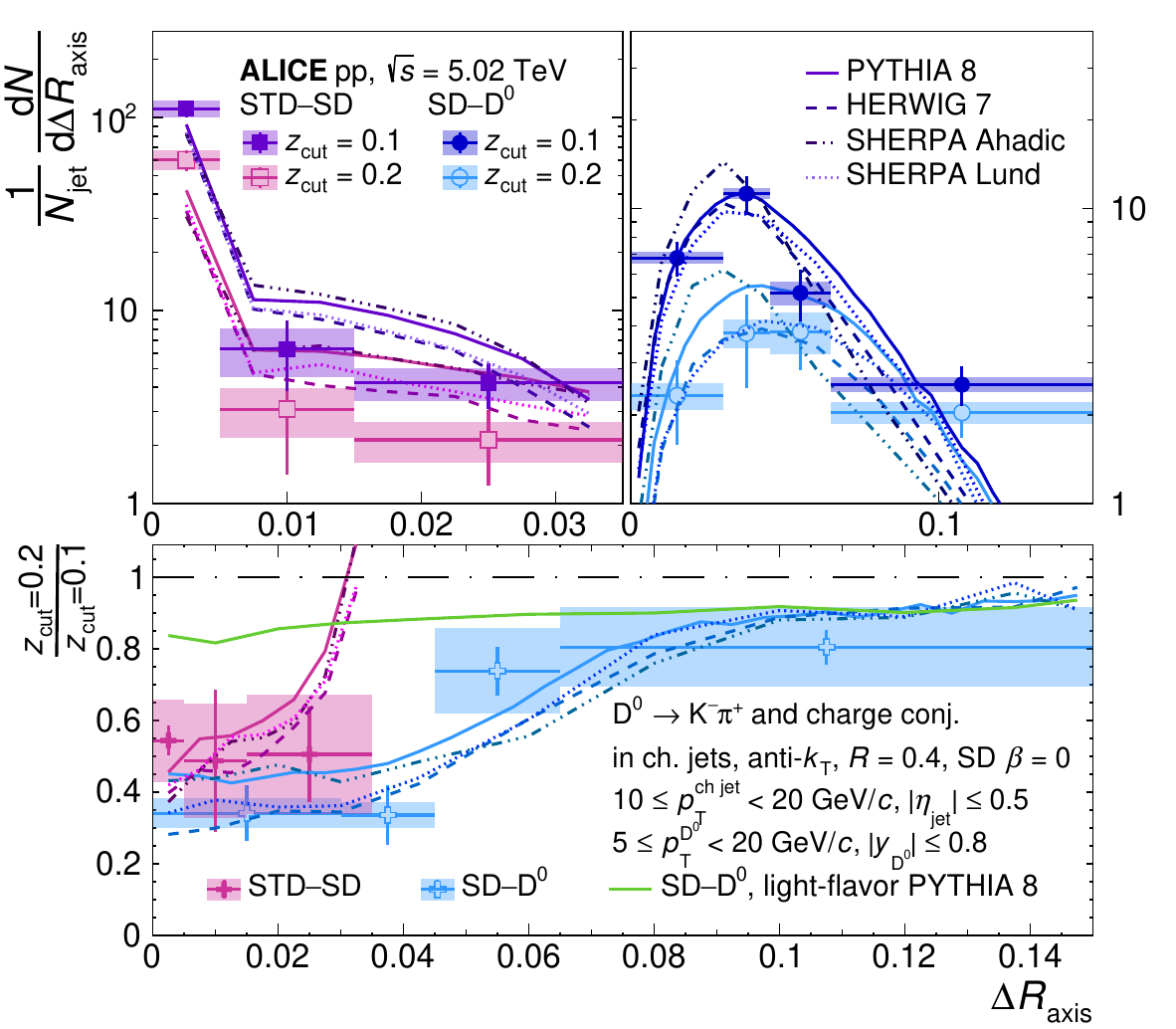}
\caption[]{Fully unfolded jet axes difference distributions for \deltaR{STD}{SD} (left) and \deltaR{SD}{\Dzero} (right), in \jetptrange{10}{20}. The Soft Drop parameter, $\zcut$, is varied between $\zcut=0.1$ and $\zcut=0.2$; the parameter $\beta=0$ throughout. Systematic and statistical uncertainties are represented by color boxes and error bars, respectively. Comparisons to MC event generators are included. The bottom panel shows a ratio of the $\zcut=0.2$ distribution over the $\zcut=0.1$ distribution for \deltaR{SD}{\Dzero} and \deltaR{STD}{SD}. These ratios are taken for both the \Dzero-tagged data and each MC event generator. }	
\label{fig:unf_zcut_ratio}
\end{figure}
%%%%%%%%%%%%%%%%%%%%%%%%%%%%%%%

To extract a more precise conclusion from the comparisons made in Fig.~\ref{fig:unf_sd-d+wta-sd_z01},~\ref{fig:unf_sd-d+wta-sd_z02}, and~\ref{fig:unf_std_sd}, we consider the ratio of the $\zcut=0.2$ distribution to $\zcut=0.1$ for both \deltaR{SD}{\Dzero} and \deltaR{STD}{SD}. This is shown in Fig.~\ref{fig:unf_zcut_ratio}. The ratio in \deltaR{STD}{SD} is, within uncertainties, flat in the region of \Rplain that is experimentally accessible, indicating that increasing the grooming intensity from $\zcut=0.1$ to $\zcut=0.2$ removes radiation uniformly with respect to the Standard axis. The p-value for a constant ratio of 0.5 is 0.99. The ratio's uniformity around 0.5 implies that the Soft Drop jet axis remains very near to the Standard jet axis and more intense grooming does not change its direction.

The ratio of the \deltaR{SD}{\Dzero} distributions shows that jets are more likely to pass the grooming condition if the Soft Drop axis is further from the \Dzero. This means that the jets surviving Soft Drop grooming are more likely to have a harder and/or wider emission that shifts the Soft Drop jet axis away from the \Dzero direction. Predictions from PYTHIA 8 of the same ratio for light-quark initiated jets show little suppression at small \Rplain compared to the \Dzero-tagged sample. This provides further evidence of the dead-cone effect~\cite{dead_cone_2022}, where harder emissions are less frequent at smaller angles from the heavy quark.

\section{Conclusions}\label{sec:conclusion}

In this work, the first measurements of the angles between the Standard, Soft Drop groomed, Winner-Takes-All and the \Dzero jet axes in jets tagged with prompt \Dzero mesons are reported. The measurement focuses on low-momentum jets, where quark-mass effects are significant, within \jetptrange{10}{50} and provides a detailed picture of the reconstructed radiation patterns of \Dzero-tagged jets. We find that the \Dzero does not always define the direction of the Standard jet axis. However, due to the strong alignment of the \Dzero and Winner-Takes-All jet axes, we confirm that the \Dzero meson is overwhelmingly the leading particle in \Dzero-tagged jets. This is observed in $(99\% \pm 1)\%$ of jets in \jetptrange{10}{20}.

The distribution of \deltaR{STD}{SD} shows that stricter grooming ($\zcut=0.2$) has minimal impact on the Soft Drop axis position with respect to the Standard axis in \Dzero-tagged jets. Increased grooming does not alter this result, as the jet axis is primarily determined by hard jet fragments. A similar observation was made for inclusive jets~\cite{Cruz_Torres_2022}. 

In contrast, the \deltaR{SD}{\Dzero} distribution shows that \Dzero-tagged jets generally have high momentum asymmetry at small \Rplain, and often fail the grooming criteria.
Because the dead cone suppresses small-angle emissions, heavy-flavor jets tend to have emissions at wider angles than inclusive jets. Those jets with a hard emission at a wider angle are more likely to survive stricter grooming, which is observed when the Soft Drop and \Dzero axes are separated.

Comparisons to several event generators show that models with string-based fragmentation reproduce both \deltaR{STD}{\Dzero} and the grooming effects on \deltaR{SD}{\Dzero}. However, the agreement with the measured \deltaR{STD}{SD} axis distribution worsens with stricter grooming, where only SHERPA Lund and HERWIG perform well. Consequently, there is not yet a clear message as to whether hadron formation entails string breaking or cluster hadronization. Comparing heavy-flavor jets to inclusive, gluon and light-flavor jets in PYTHIA 8 shows the impact of flavor effects on the jet axis, and the contribution of single-track jets is clearly visualized as a consequence of the dead-cone effect.

In future analyses, the inclusion of the $\Lambda_{\rm c}^{+}$ baryon with a measurement of the $\rm \Lambda_c^+ / D^0$ ratio may provide additional insight into charm hadronization. These results in \pp collision data lay the groundwork for studies of \Dzero-tagged jets in Pb--Pb collisions, which may be sensitive to the diffusion properties of the QGP and could provide pivotal insights into charm-quark energy loss. Additionally, future studies may elucidate the specific impact of the QGP on heavy-flavor quark fragmentation and radiation patterns, contributing to a deeper understanding of QCD dynamics in dense nuclear environments.

%%%%%%%%%%%%%%%%%%%%%%%%%%%%%%%%
% end main text 
%%%%%%%%%%%%%%%%%%%%%%%%%%%%%%%%

%%%%% acknowledgements - handled by EB chairs 
\newenvironment{acknowledgement}{\relax}{\relax}
\begin{acknowledgement}
\section*{Acknowledgements}
% add specific acknowledgements here 
% ...but please don't remove the line below: funding agencies
% will be acknowledged with a custom tex file handled by EB chairs after Collab Round 2
% Version: 2025-03-21

The ALICE Collaboration would like to thank all its engineers and technicians for their invaluable contributions to the construction of the experiment and the CERN accelerator teams for the outstanding performance of the LHC complex.
The ALICE Collaboration gratefully acknowledges the resources and support provided by all Grid centres and the Worldwide LHC Computing Grid (WLCG) collaboration.
The ALICE Collaboration acknowledges the following funding agencies for their support in building and running the ALICE detector:
A. I. Alikhanyan National Science Laboratory (Yerevan Physics Institute) Foundation (ANSL), State Committee of Science and World Federation of Scientists (WFS), Armenia;
Austrian Academy of Sciences, Austrian Science Fund (FWF): [M 2467-N36] and Nationalstiftung f\"{u}r Forschung, Technologie und Entwicklung, Austria;
Ministry of Communications and High Technologies, National Nuclear Research Center, Azerbaijan;
Conselho Nacional de Desenvolvimento Cient\'{\i}fico e Tecnol\'{o}gico (CNPq), Financiadora de Estudos e Projetos (Finep), Funda\c{c}\~{a}o de Amparo \`{a} Pesquisa do Estado de S\~{a}o Paulo (FAPESP) and Universidade Federal do Rio Grande do Sul (UFRGS), Brazil;
Bulgarian Ministry of Education and Science, within the National Roadmap for Research Infrastructures 2020-2027 (object CERN), Bulgaria;
Ministry of Education of China (MOEC) , Ministry of Science \& Technology of China (MSTC) and National Natural Science Foundation of China (NSFC), China;
Ministry of Science and Education and Croatian Science Foundation, Croatia;
Centro de Aplicaciones Tecnol\'{o}gicas y Desarrollo Nuclear (CEADEN), Cubaenerg\'{\i}a, Cuba;
Ministry of Education, Youth and Sports of the Czech Republic, Czech Republic;
The Danish Council for Independent Research | Natural Sciences, the VILLUM FONDEN and Danish National Research Foundation (DNRF), Denmark;
Helsinki Institute of Physics (HIP), Finland;
Commissariat \`{a} l'Energie Atomique (CEA) and Institut National de Physique Nucl\'{e}aire et de Physique des Particules (IN2P3) and Centre National de la Recherche Scientifique (CNRS), France;
Bundesministerium f\"{u}r Bildung und Forschung (BMBF) and GSI Helmholtzzentrum f\"{u}r Schwerionenforschung GmbH, Germany;
General Secretariat for Research and Technology, Ministry of Education, Research and Religions, Greece;
National Research, Development and Innovation Office, Hungary;
Department of Atomic Energy Government of India (DAE), Department of Science and Technology, Government of India (DST), University Grants Commission, Government of India (UGC) and Council of Scientific and Industrial Research (CSIR), India;
National Research and Innovation Agency - BRIN, Indonesia;
Istituto Nazionale di Fisica Nucleare (INFN), Italy;
Japanese Ministry of Education, Culture, Sports, Science and Technology (MEXT) and Japan Society for the Promotion of Science (JSPS) KAKENHI, Japan;
Consejo Nacional de Ciencia (CONACYT) y Tecnolog\'{i}a, through Fondo de Cooperaci\'{o}n Internacional en Ciencia y Tecnolog\'{i}a (FONCICYT) and Direcci\'{o}n General de Asuntos del Personal Academico (DGAPA), Mexico;
Nederlandse Organisatie voor Wetenschappelijk Onderzoek (NWO), Netherlands;
The Research Council of Norway, Norway;
Pontificia Universidad Cat\'{o}lica del Per\'{u}, Peru;
Ministry of Science and Higher Education, National Science Centre and WUT ID-UB, Poland;
Korea Institute of Science and Technology Information and National Research Foundation of Korea (NRF), Republic of Korea;
Ministry of Education and Scientific Research, Institute of Atomic Physics, Ministry of Research and Innovation and Institute of Atomic Physics and Universitatea Nationala de Stiinta si Tehnologie Politehnica Bucuresti, Romania;
Ministerstvo skolstva, vyskumu, vyvoja a mladeze SR, Slovakia;
National Research Foundation of South Africa, South Africa;
Swedish Research Council (VR) and Knut \& Alice Wallenberg Foundation (KAW), Sweden;
European Organization for Nuclear Research, Switzerland;
Suranaree University of Technology (SUT), National Science and Technology Development Agency (NSTDA) and National Science, Research and Innovation Fund (NSRF via PMU-B B05F650021), Thailand;
Turkish Energy, Nuclear and Mineral Research Agency (TENMAK), Turkey;
National Academy of  Sciences of Ukraine, Ukraine;
Science and Technology Facilities Council (STFC), United Kingdom;
National Science Foundation of the United States of America (NSF) and United States Department of Energy, Office of Nuclear Physics (DOE NP), United States of America.
In addition, individual groups or members have received support from:
Czech Science Foundation (grant no. 23-07499S), Czech Republic;
FORTE project, reg.\ no.\ CZ.02.01.01/00/22\_008/0004632, Czech Republic, co-funded by the European Union, Czech Republic;
European Research Council (grant no. 950692), European Union;
Deutsche Forschungs Gemeinschaft (DFG, German Research Foundation) ``Neutrinos and Dark Matter in Astro- and Particle Physics'' (grant no. SFB 1258), Germany;
ICSC - National Research Center for High Performance Computing, Big Data and Quantum Computing and FAIR - Future Artificial Intelligence Research, funded by the NextGenerationEU program (Italy).

\end{acknowledgement}

%%%%%%%% Bibliography 
\bibliographystyle{utphys}   % Remember we use title in the biblio
\bibliography{bib/ref_introduction,bib/ref_experimentalsetup,bib/ref_datasampleandselection,bib/ref_d0reconstructionandselections,bib/ref_corrections}

%%%%%%%%%%%%%%%%%%%%%%%%%%%%%%%%
% Appendices: yours (if any) + authorlist
%%%%%%%%%%%%%%%%%%%%%%%%%%%%%%%%
\newpage
\appendix

%
%\input{} % put your appendices here (if any)
%

%%%%% Authorlist - please do not touch: handled by EB chairs 
\section{The ALICE Collaboration}
\label{app:collab}
% ALICE Collaboration author list for 2025-03-21
\begin{flushleft} 
\small

S.~Acharya\,\orcidlink{0000-0002-9213-5329}\,$^{\rm 50}$, 
A.~Agarwal$^{\rm 133}$, 
G.~Aglieri Rinella\,\orcidlink{0000-0002-9611-3696}\,$^{\rm 32}$, 
L.~Aglietta\,\orcidlink{0009-0003-0763-6802}\,$^{\rm 24}$, 
M.~Agnello\,\orcidlink{0000-0002-0760-5075}\,$^{\rm 29}$, 
N.~Agrawal\,\orcidlink{0000-0003-0348-9836}\,$^{\rm 25}$, 
Z.~Ahammed\,\orcidlink{0000-0001-5241-7412}\,$^{\rm 133}$, 
S.~Ahmad\,\orcidlink{0000-0003-0497-5705}\,$^{\rm 15}$, 
S.U.~Ahn\,\orcidlink{0000-0001-8847-489X}\,$^{\rm 71}$, 
I.~Ahuja\,\orcidlink{0000-0002-4417-1392}\,$^{\rm 36}$
% ...

\section*{Affiliation Notes}

$^{\rm I}$ Deceased\\
$^{\rm II}$ Also at: Max-Planck-Institut fur Physik, Munich, Germany\\
$^{\rm III}$ Also at: Italian National Agency for New Technologies, Energy and Sustainable Economic Development (ENEA), Bologna, Italy\\
$^{\rm IV}$ Also at: Instituto de Fisica da Universidade de Sao Paulo\\
$^{\rm V}$ Also at: Dipartimento DET del Politecnico di Torino, Turin, Italy\\
$^{\rm VI}$ Also at: Department of Applied Physics, Aligarh Muslim University, Aligarh, India\\
$^{\rm VII}$ Also at: Institute of Theoretical Physics, University of Wroclaw, Poland\\
$^{\rm VIII}$ Also at: Facultad de Ciencias, Universidad Nacional Aut\'{o}noma de M\'{e}xico, Mexico City, Mexico\\

\section*{Collaboration Institutes}

$^{1}$ A.I. Alikhanyan National Science Laboratory (Yerevan Physics Institute) Foundation, Yerevan, Armenia\\
$^{2}$ AGH University of Krakow, Cracow, Poland\\
$^{3}$ Bogolyubov Institute for Theoretical Physics, National Academy of Sciences of Ukraine, Kiev, Ukraine\\
$^{4}$ Bose Institute, Department of Physics  and Centre for Astroparticle Physics and Space Science (CAPSS), Kolkata, India\\
$^{5}$ California Polytechnic State University, San Luis Obispo, California, United States\\
$^{6}$ Central China Normal University, Wuhan, China\\
$^{7}$ Centro de Aplicaciones Tecnol\'{o}gicas y Desarrollo Nuclear (CEADEN), Havana, Cuba\\
$^{8}$ Centro de Investigaci\'{o}n y de Estudios Avanzados (CINVESTAV), Mexico City and M\'{e}rida, Mexico\\
$^{9}$ Chicago State University, Chicago, Illinois, United States\\
$^{10}$ China Nuclear Data Center, China Institute of Atomic Energy, Beijing, China\\
$^{11}$ China University of Geosciences, Wuhan, China\\
$^{12}$ Chungbuk National University, Cheongju, Republic of Korea\\
$^{13}$ Comenius University Bratislava, Faculty of Mathematics, Physics and Informatics, Bratislava, Slovak Republic\\
$^{14}$ Creighton University, Omaha, Nebraska, United States\\
$^{15}$ Department of Physics, Aligarh Muslim University, Aligarh, India\\
$^{16}$ Department of Physics, Pusan National University, Pusan, Republic of Korea\\
$^{17}$ Department of Physics, Sejong University, Seoul, Republic of Korea\\
$^{18}$ Department of Physics, University of California, Berkeley, California, United States\\
$^{19}$ Department of Physics, University of Oslo, Oslo, Norway\\
$^{20}$ Department of Physics and Technology, University of Bergen, Bergen, Norway\\
$^{21}$ Dipartimento di Fisica, Universit\`{a} di Pavia, Pavia, Italy\\
$^{22}$ Dipartimento di Fisica dell'Universit\`{a} and Sezione INFN, Cagliari, Italy\\
$^{23}$ Dipartimento di Fisica dell'Universit\`{a} and Sezione INFN, Trieste, Italy\\
$^{24}$ Dipartimento di Fisica dell'Universit\`{a} and Sezione INFN, Turin, Italy\\
$^{25}$ Dipartimento di Fisica e Astronomia dell'Universit\`{a} and Sezione INFN, Bologna, Italy\\
$^{26}$ Dipartimento di Fisica e Astronomia dell'Universit\`{a} and Sezione INFN, Catania, Italy\\
$^{27}$ Dipartimento di Fisica e Astronomia dell'Universit\`{a} and Sezione INFN, Padova, Italy\\
$^{28}$ Dipartimento di Fisica `E.R.~Caianiello' dell'Universit\`{a} and Gruppo Collegato INFN, Salerno, Italy\\
$^{29}$ Dipartimento DISAT del Politecnico and Sezione INFN, Turin, Italy\\
$^{30}$ Dipartimento di Scienze MIFT, Universit\`{a} di Messina, Messina, Italy\\
$^{31}$ Dipartimento Interateneo di Fisica `M.~Merlin' and Sezione INFN, Bari, Italy\\
$^{32}$ European Organization for Nuclear Research (CERN), Geneva, Switzerland\\
$^{33}$ Faculty of Electrical Engineering, Mechanical Engineering and Naval Architecture, University of Split, Split, Croatia\\
$^{34}$ Faculty of Nuclear Sciences and Physical Engineering, Czech Technical University in Prague, Prague, Czech Republic\\
$^{35}$ Faculty of Physics, Sofia University, Sofia, Bulgaria\\
$^{36}$ Faculty of Science, P.J.~\v{S}af\'{a}rik University, Ko\v{s}ice, Slovak Republic\\
$^{37}$ Faculty of Technology, Environmental and Social Sciences, Bergen, Norway\\
$^{38}$ Frankfurt Institute for Advanced Studies, Johann Wolfgang Goethe-Universit\"{a}t Frankfurt, Frankfurt, Germany\\
$^{39}$ Fudan University, Shanghai, China\\
$^{40}$ Gangneung-Wonju National University, Gangneung, Republic of Korea\\
$^{41}$ Gauhati University, Department of Physics, Guwahati, India\\
$^{42}$ Helmholtz-Institut f\"{u}r Strahlen- und Kernphysik, Rheinische Friedrich-Wilhelms-Universit\"{a}t Bonn, Bonn, Germany\\
$^{43}$ Helsinki Institute of Physics (HIP), Helsinki, Finland\\
$^{44}$ High Energy Physics Group,  Universidad Aut\'{o}noma de Puebla, Puebla, Mexico\\
$^{45}$ Horia Hulubei National Institute of Physics and Nuclear Engineering, Bucharest, Romania\\
$^{46}$ HUN-REN Wigner Research Centre for Physics, Budapest, Hungary\\
$^{47}$ Indian Institute of Technology Bombay (IIT), Mumbai, India\\
$^{48}$ Indian Institute of Technology Indore, Indore, India\\
$^{49}$ INFN, Laboratori Nazionali di Frascati, Frascati, Italy\\
$^{50}$ INFN, Sezione di Bari, Bari, Italy\\
$^{51}$ INFN, Sezione di Bologna, Bologna, Italy\\
$^{52}$ INFN, Sezione di Cagliari, Cagliari, Italy\\
$^{53}$ INFN, Sezione di Catania, Catania, Italy\\
$^{54}$ INFN, Sezione di Padova, Padova, Italy\\
$^{55}$ INFN, Sezione di Pavia, Pavia, Italy\\
$^{56}$ INFN, Sezione di Torino, Turin, Italy\\
$^{57}$ INFN, Sezione di Trieste, Trieste, Italy\\
$^{58}$ Inha University, Incheon, Republic of Korea\\
$^{59}$ Institute for Gravitational and Subatomic Physics (GRASP), Utrecht University/Nikhef, Utrecht, Netherlands\\
$^{60}$ Institute of Experimental Physics, Slovak Academy of Sciences, Ko\v{s}ice, Slovak Republic\\
$^{61}$ Institute of Physics, Homi Bhabha National Institute, Bhubaneswar, India\\
$^{62}$ Institute of Physics of the Czech Academy of Sciences, Prague, Czech Republic\\
$^{63}$ Institute of Space Science (ISS), Bucharest, Romania\\
$^{64}$ Institut f\"{u}r Kernphysik, Johann Wolfgang Goethe-Universit\"{a}t Frankfurt, Frankfurt, Germany\\
$^{65}$ Instituto de Ciencias Nucleares, Universidad Nacional Aut\'{o}noma de M\'{e}xico, Mexico City, Mexico\\
$^{66}$ Instituto de F\'{i}sica, Universidade Federal do Rio Grande do Sul (UFRGS), Porto Alegre, Brazil\\
$^{67}$ Instituto de F\'{\i}sica, Universidad Nacional Aut\'{o}noma de M\'{e}xico, Mexico City, Mexico\\
$^{68}$ iThemba LABS, National Research Foundation, Somerset West, South Africa\\
$^{69}$ Jeonbuk National University, Jeonju, Republic of Korea\\
$^{70}$ Johann-Wolfgang-Goethe Universit\"{a}t Frankfurt Institut f\"{u}r Informatik, Fachbereich Informatik und Mathematik, Frankfurt, Germany\\
$^{71}$ Korea Institute of Science and Technology Information, Daejeon, Republic of Korea\\
$^{72}$ Laboratoire de Physique Subatomique et de Cosmologie, Universit\'{e} Grenoble-Alpes, CNRS-IN2P3, Grenoble, France\\
$^{73}$ Lawrence Berkeley National Laboratory, Berkeley, California, United States\\
$^{74}$ Lund University Department of Physics, Division of Particle Physics, Lund, Sweden\\
$^{75}$ Nagasaki Institute of Applied Science, Nagasaki, Japan\\
$^{76}$ Nara Women{'}s University (NWU), Nara, Japan\\
$^{77}$ National and Kapodistrian University of Athens, School of Science, Department of Physics , Athens, Greece\\
$^{78}$ National Centre for Nuclear Research, Warsaw, Poland\\
$^{79}$ National Institute of Science Education and Research, Homi Bhabha National Institute, Jatni, India\\
$^{80}$ National Nuclear Research Center, Baku, Azerbaijan\\
$^{81}$ National Research and Innovation Agency - BRIN, Jakarta, Indonesia\\
$^{82}$ Niels Bohr Institute, University of Copenhagen, Copenhagen, Denmark\\
$^{83}$ Nikhef, National institute for subatomic physics, Amsterdam, Netherlands\\
$^{84}$ Nuclear Physics Group, STFC Daresbury Laboratory, Daresbury, United Kingdom\\
$^{85}$ Nuclear Physics Institute of the Czech Academy of Sciences, Husinec-\v{R}e\v{z}, Czech Republic\\
$^{86}$ Oak Ridge National Laboratory, Oak Ridge, Tennessee, United States\\
$^{87}$ Ohio State University, Columbus, Ohio, United States\\
$^{88}$ Physics department, Faculty of science, University of Zagreb, Zagreb, Croatia\\
$^{89}$ Physics Department, Panjab University, Chandigarh, India\\
$^{90}$ Physics Department, University of Jammu, Jammu, India\\
$^{91}$ Physics Program and International Institute for Sustainability with Knotted Chiral Meta Matter (WPI-SKCM$^{2}$), Hiroshima University, Hiroshima, Japan\\
$^{92}$ Physikalisches Institut, Eberhard-Karls-Universit\"{a}t T\"{u}bingen, T\"{u}bingen, Germany\\
$^{93}$ Physikalisches Institut, Ruprecht-Karls-Universit\"{a}t Heidelberg, Heidelberg, Germany\\
$^{94}$ Physik Department, Technische Universit\"{a}t M\"{u}nchen, Munich, Germany\\
$^{95}$ Politecnico di Bari and Sezione INFN, Bari, Italy\\
$^{96}$ Research Division and ExtreMe Matter Institute EMMI, GSI Helmholtzzentrum f\"ur Schwerionenforschung GmbH, Darmstadt, Germany\\
$^{97}$ Saga University, Saga, Japan\\
$^{98}$ Saha Institute of Nuclear Physics, Homi Bhabha National Institute, Kolkata, India\\
$^{99}$ School of Physics and Astronomy, University of Birmingham, Birmingham, United Kingdom\\
$^{100}$ Secci\'{o}n F\'{\i}sica, Departamento de Ciencias, Pontificia Universidad Cat\'{o}lica del Per\'{u}, Lima, Peru\\
$^{101}$ Stefan Meyer Institut f\"{u}r Subatomare Physik (SMI), Vienna, Austria\\
$^{102}$ SUBATECH, IMT Atlantique, Nantes Universit\'{e}, CNRS-IN2P3, Nantes, France\\
$^{103}$ Sungkyunkwan University, Suwon City, Republic of Korea\\
$^{104}$ Suranaree University of Technology, Nakhon Ratchasima, Thailand\\
$^{105}$ Technical University of Ko\v{s}ice, Ko\v{s}ice, Slovak Republic\\
$^{106}$ The Henryk Niewodniczanski Institute of Nuclear Physics, Polish Academy of Sciences, Cracow, Poland\\
$^{107}$ The University of Texas at Austin, Austin, Texas, United States\\
$^{108}$ Universidad Aut\'{o}noma de Sinaloa, Culiac\'{a}n, Mexico\\
$^{109}$ Universidade de S\~{a}o Paulo (USP), S\~{a}o Paulo, Brazil\\
$^{110}$ Universidade Estadual de Campinas (UNICAMP), Campinas, Brazil\\
$^{111}$ Universidade Federal do ABC, Santo Andre, Brazil\\
$^{112}$ Universitatea Nationala de Stiinta si Tehnologie Politehnica Bucuresti, Bucharest, Romania\\
$^{113}$ University of Derby, Derby, United Kingdom\\
$^{114}$ University of Houston, Houston, Texas, United States\\
$^{115}$ University of Jyv\"{a}skyl\"{a}, Jyv\"{a}skyl\"{a}, Finland\\
$^{116}$ University of Kansas, Lawrence, Kansas, United States\\
$^{117}$ University of Liverpool, Liverpool, United Kingdom\\
$^{118}$ University of Science and Technology of China, Hefei, China\\
$^{119}$ University of South-Eastern Norway, Kongsberg, Norway\\
$^{120}$ University of Tennessee, Knoxville, Tennessee, United States\\
$^{121}$ University of the Witwatersrand, Johannesburg, South Africa\\
$^{122}$ University of Tokyo, Tokyo, Japan\\
$^{123}$ University of Tsukuba, Tsukuba, Japan\\
$^{124}$ Universit\"{a}t M\"{u}nster, Institut f\"{u}r Kernphysik, M\"{u}nster, Germany\\
$^{125}$ Universit\'{e} Clermont Auvergne, CNRS/IN2P3, LPC, Clermont-Ferrand, France\\
$^{126}$ Universit\'{e} de Lyon, CNRS/IN2P3, Institut de Physique des 2 Infinis de Lyon, Lyon, France\\
$^{127}$ Universit\'{e} de Strasbourg, CNRS, IPHC UMR 7178, F-67000 Strasbourg, France, Strasbourg, France\\
$^{128}$ Universit\'{e} Paris-Saclay, Centre d'Etudes de Saclay (CEA), IRFU, D\'{e}partment de Physique Nucl\'{e}aire (DPhN), Saclay, France\\
$^{129}$ Universit\'{e}  Paris-Saclay, CNRS/IN2P3, IJCLab, Orsay, France\\
$^{130}$ Universit\`{a} degli Studi di Foggia, Foggia, Italy\\
$^{131}$ Universit\`{a} del Piemonte Orientale, Vercelli, Italy\\
$^{132}$ Universit\`{a} di Brescia, Brescia, Italy\\
$^{133}$ Variable Energy Cyclotron Centre, Homi Bhabha National Institute, Kolkata, India\\
$^{134}$ Warsaw University of Technology, Warsaw, Poland\\
$^{135}$ Wayne State University, Detroit, Michigan, United States\\
$^{136}$ Yale University, New Haven, Connecticut, United States\\
$^{137}$ Yildiz Technical University, Istanbul, Turkey\\
$^{138}$ Yonsei University, Seoul, Republic of Korea\\
$^{139}$ Affiliated with an institute formerly covered by a cooperation agreement with CERN\\
$^{140}$ Affiliated with an international laboratory covered by a cooperation agreement with CERN.\\

\end{flushleft} 
  
\end{document}